\documentclass{aa}  

\usepackage{natbib}
\usepackage{graphicx}
\usepackage{caption}
\usepackage{subcaption}
\usepackage{lipsum}
\usepackage{inputenc}
\usepackage{color}
\usepackage{amsmath,amsfonts,amssymb}
\usepackage{float}
\usepackage{hyperref}
\usepackage[version=3]{mhchem}
\usepackage{lineno}
\usepackage{amsmath}

\begin{document}

   \title{Mapping the zonal winds of Jupiter's stratospheric equatorial oscillation}

   \author{B. Benmahi
          \inst{1}
          \and
          T. Cavali\'e\inst{1,2}
          \and
          T. K. Greathouse\inst{3}
          \and
          V. Hue\inst{3}       
          \and
          R. Giles\inst{3}
          \and
          S. Guerlet\inst{4}
          \and
          A. Spiga\inst{4, 5}
          \and
          R. Cosentino\inst{6}}
          
   \institute{Laboratoire d'Astrophysique de Bordeaux, Univ. Bordeaux, CNRS, B18N, all\'ee Geoffroy Saint-Hilaire, 33615 Pessac, France.\\
              \email{bilal.benmahi$@$u-bordeaux.fr}
         \and
            LESIA, Observatoire de Paris, Universit\'e PSL, CNRS, Sorbonne Universit\'e, Univ. Paris Diderot, Sorbonne Paris Cit\'e, 5 place Jules Janssen, 92195 Meudon, France.
        \and
            Southwest Research Institute, San Antonio, TX 78228, USA.
        \and
        	Laboratoire de Météorologie Dynamique / Institut Pierre-Simon Laplace (LMD/IPSL), Sorbonne Université, Centre National de la Recherche Scientifique (CNRS), Ecole Polytechnique, Ecole Normale Supérieure (ENS), address: Campus Pierre et Marie Curie BC99, 4 place Jussieu, 75005 Paris, France.
        \and Institut Universitaire de France, Paris.
        \and
        	Department of Astronomy, University of Maryland, College Park, MD 20742, USA.}

   \date{Received 11 june 2021 ; Accepted 12 july 2021}

 
  \abstract
   {
   	Since the 1950s, quasi-periodic oscillations have been studied in the terrestrial equatorial stratosphere. Other planets of the solar system present (or are expected to present) such oscillations, like the Jupiter Equatorial Oscillation (JEO) and the Saturn Semi-Annual Oscillation (SSAO).  
   	In Jupiter's stratosphere, the equatorial oscillation of its relative temperature structure about the equator, is characterized by a quasi-period of 4.4 years.
    }
   { 
   	The stratospheric wind field in Jupiter's equatorial zone has never been directly observed. 
	In this paper, we aim at mapping the absolute wind speeds in Jupiter's equatorial stratosphere to quantify vertical and horizontal wind and temperature shear.
	}
   {
   	Assuming geostrophic equilibrium, we apply the thermal wind balance using nearly simultaneous stratospheric temperature measurements between 0.1 and 30\,mbar performed with Gemini/TEXES and direct zonal wind measurements derived at 1\,mbar from ALMA observations, all carried out between March 14$^\mathrm{th}$ and 22$^\mathrm{nd}$, 2017. We are thus able to calculate self-consistently the zonal wind field in Jupiter's stratosphere where the JEO occurs.
   	
 }
   {
   	We obtain stratospheric map of the zonal wind speeds as a function of latitude and pressure about Jupiter's equator for the first time. The winds are vertically layered with successive eastward and westward jets. We find a 200\,m/s westward jet at 4\,mbar at the equator, with a typical longitudinal variability on the order of $\sim$50\,m/s.
   	By extending our wind calculations to the upper troposphere, we find a wind structure qualitatively close to the wind observed using cloud-tracking techniques.
   }
   { 
   	Nearly simultaneous temperature and wind measurements, both in the stratosphere, are a powerful tool for future investigations of the JEO (and other planetary equatorial oscillations) and its temporal evolution.
   	
   	
   }

   \keywords{Planets and satellites: individual: Jupiter ; Planets and satellites: atmospheres}
   \authorrunning{B. Benmahi et al.}
   \maketitle
%
\section{Introduction}
\label{section:intro}
In Earth's atmosphere, \cite{reed_evidence_1961} and \cite{ebdon_fluctuations_1961} discovered a quasi-periodic oscillation in the high-altitude equatorial winds, alternating between eastward and westward flows. This $\sim$ 28 months quasi-periodic phenomenon, known as the quasi-biennial oscillation (QBO) has been observed and studied ever since the 1950's (e.g. \citealt{baldwin_quasi-biennial_2001}). The thermal wind balance indicates that atmospheric winds are coupled with temperature gradients. The QBO can then also be characterized from a temperature field standpoint. From this perspective, the QBO is characterized by a vertical oscillation in temperature which moves downwards in the stratosphere with time. 

Using temperature field measurements using infrared observations, similar quasi-periodic stratospheric oscillations have been discovered in Jupiter \citep{leovy_quasiquadrennial_1991,orton_thermal_1991} and Saturn \citep{fouchet_equatorial_2008,orton_semi-annual_2008} and are all localized in the 20$^{\circ}$S - 20$^{\circ}$N latitudinal range. In Jupiter, the oscillation is characterized by a period of $\sim$ 4-years and was dubbed the quasi-quadrennial oscillation (QQO). Further temperature observations have enabled its characterization and modeling (e.g., \citealt{orton_thermal_1991,friedson_new_1999,simon-miller_jupiters_2006,fletcher_mid-infrared_2016}). The Saturn Semi-Annual Oscillation (SSAO) has a period of about 14.7 years. Both have been observed between 0.01 and 20\,mbar. 

Planetary and gravity waves were first proposed as the cause of the Earth QBO by \citet{lindzen_theory_1968} and have since been proposed as causes of the Jupiter and Saturn oscillations as well \citep{friedson_new_1999,Li2000,flasar_intense_2004,cosentino_new_2017,bardet_global_2021}. In the Earth, momentum transfer from the waves to the zonal wind results in downward propagation of the wind velocity peaks. Such downward propagation of the SSAO was observed, with the Composite InfraRed Spectrometer (CIRS), over the 13-year course of the Cassini mission \citep{guerlet_evolution_2011,guerlet_equatorial_2018}. The downward propagation of the JEO has also been measured from long-term monitoring temperature observations carried out at the NASA Infrared Telescope Facility (IRTF) using the Texas Echelon Cross-Echelle Spectrograph (TEXES) \citep{cosentino_new_2017,cosentino_effects_2020,giles_vertically-resolved_2020}, which allow the retrieval of horizontally and vertically resolved stratospheric temperatures. Recently, \cite{antunano_long-term_2020} showed that Jupiter's QQO does not have a stable periodicity, and alterations could result from thermal perturbations \citep{giles_vertically-resolved_2020}. We will therefore refer to the Jupiter Equatorial Oscillation (JEO) rather than the QQO in what follows. We note that such perturbations have also been seen in the SSAO following the Great Storm of 2010-2011 \citep{Fletcher2017}. 

\cite{flasar_intense_2004} derived the JEO wind structure at low latitudes from Cassini/CIRS temperature measurements performed during the Jupiter flyby in late 2000. Invoking thermal wind balance, they discovered a $\sim$ 140\,m/s eastward equatorial jet at 3\,mbar by interpolating wind velocities, even though a precise estimate of the jet peak velocity at the equator is made impossible because of the increase of the Rossby number. \citet{fletcher_mid-infrared_2016} confirmed the presence of a strong stratospheric jet in the\,mbar region. \cite{cosentino_new_2017} were able to reproduce such JEO wind speeds, to first order, using an atmospheric circulation model with a stochastic parametrization of gravity wave drag. 

In Saturn's stratosphere, \citet{fouchet_equatorial_2008} found a velocity difference of about 200\,m/s between the two equatorial jets located at 0.3 and 3\,mbar also applying the thermal wind balance to the measured temperatures. However, the absolute stratospheric wind speeds remain unknown on Saturn (and were unknown until now for Jupiter) and, given the strong eastward cloud-top zonal wind of 400\,m/s observed at $\sim$700\,mbar, it is unsure whether the stratospheric wind remains eastward or becomes periodically westward, as observed on Earth.

The main weakness in the methodology used in previous studies to derive the thermal winds and particularly in the Jupiter's case, is that there is a discontinuity between the pressure range probed by the temperatures (1-20\,mbar from \ce{CH4} emission, 80-400\,mbar from \ce{H2} collision-induced absorption) and the pressure at which the zonal wind profile is inserted as a boundary condition, i.e. generally at the cloud-top (at 500\,mbar) during the thermal wind derivation. The novelty of the approach we present here consists in using a wind measurement performed almost concomitantly and within the altitude range probed by the stratospheric temperature measurements. We are thus able to obtain self-consistent zonal wind field as a function of altitude and latitude in the whole range probed by the temperature measurements. In this paper, we focus on Jupiter's zonal winds in the altitude and latitude ranges where the JEO takes place to constrain the direction and magnitude of the equatorial and tropical jets using the thermal wind balance.

In section \ref{section:obs}, we present the wind and temperature observations we used to compute the zonal wind field. Section\ref{section:models} details the models we developed to compute the equatorial and tropical wind speeds from the thermal wind balance. We present our results and discuss them in Section \ref{section:results}.
We give our concluding remarks in Section \ref{section:conclusion}.

\section{Observations}
\label{section:obs}

\subsection{Zonal wind measurements at 1\,mbar in Jupiter's stratosphere}
\begin{figure}[h]
	\centering
	\includegraphics[width=9cm, keepaspectratio]{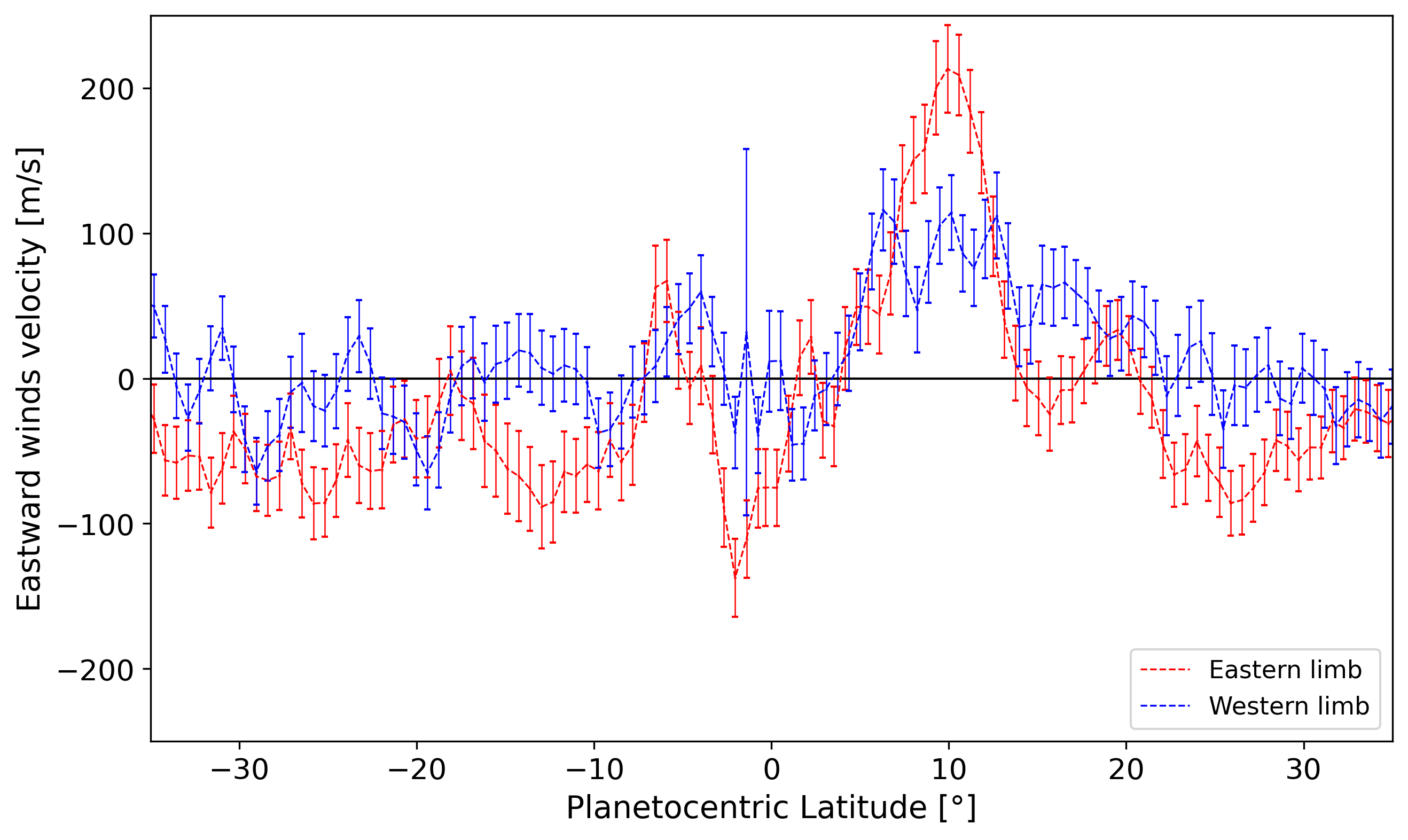}
	\caption{Eastward wind velocities at 1\,mbar as measured with ALMA on 22 March 2017, on the eastern and western limbs of Jupiter (adapted from \citealt{cavalie_first_2021}).
	}
	\label{fig:wind_east-west}
\end{figure}

\begin{figure}[h]
	\centering
	\includegraphics[width=9cm, keepaspectratio]{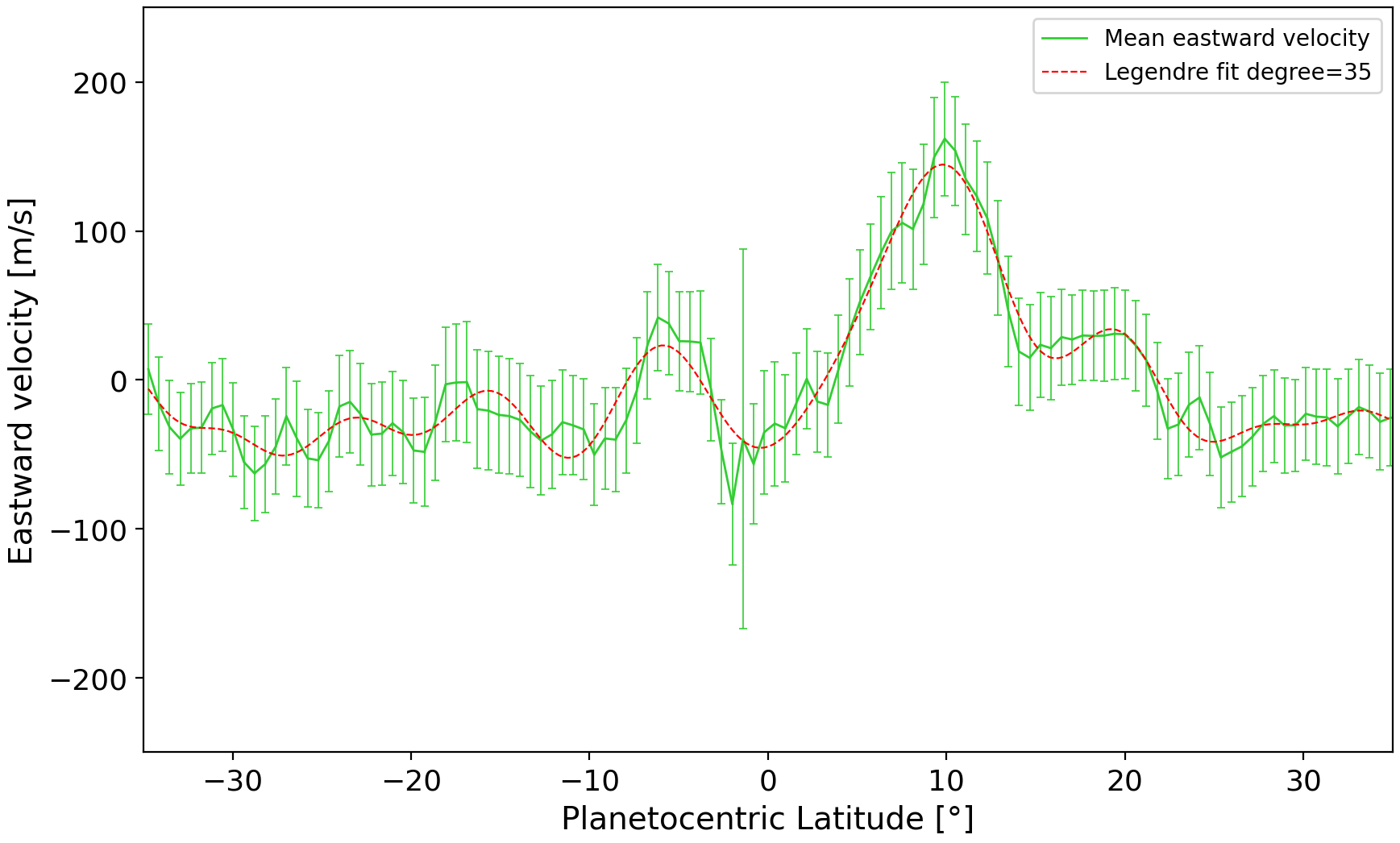}
	\caption{Mean eastward wind velocities (green line) with uncertainties (green bars) at 1\,mbar in Jupiter's stratosphere on 22 March 2017. A degree 35 Legendre polynomial smoothing is plotted with dashed red lines.
	}
	\label{fig:wind_fited}
\end{figure}

\cite{cavalie_first_2021} observed Jupiter's stratospheric HCN emission at 354.505 GHz with the ALMA interferometer on 22 March 2017. They obtained a high spectral and spatial resolution map of Jupiter's limb from which they achieved the first direct measurement of the stratospheric winds. The wind speeds were retrieved from the wind-induced Doppler-shifted spectral lines formed at the altitude probed by the HCN line. The latitudinal resolution varies from ~3$^{\circ}$ at the equator to ~7$^{\circ}$ at polar latitudes. Contribution function computations demonstrate that the sensitivity to winds peaks at 1\,mbar in the 60$^{\circ}$S-50$^{\circ}$N latitudinal range, and at 0.1\,mbar at polar latitudes. In this paper, we use the data ranging from 35$^{\circ}$S to 35$^{\circ}$N planetocentric latitude.
Although the data were acquired with a short 24-min on-source integration time, the rapid rotation of the planet (9 hrs 56 min) results in longitudinal smearing over about 15$^{\circ}$. The central meridian longitude thus ranges from 65$^{\circ}$W to 80$^{\circ}$W (System III). The eastern and western limbs (From the observer's point of view) span longitudes from 335$^{\circ}$W to 350$^{\circ}$W and from 155$^{\circ}$W to 170$^{\circ}$W respectively. 
The eastward wind velocities obtained by \cite{cavalie_first_2021} are shown for both observed limbs in Figure \ref{fig:wind_east-west}. The average of both limb measurements are shown in Figure \ref{fig:wind_fited}.

\subsection{Jupiter's stratospheric temperature field observations}

In March 2017, TEXES \citep{lacy_texes_2002}, mounted on the Gemini North 8-m telescope, carried out high-resolution infrared observations of Jupiter to characterize the temperatures in its stratosphere. These observations were taken as part of a long-term monitoring program carried out primarily at the NASA Infrared Telescope Facility \citep{cosentino_new_2017,giles_vertically-resolved_2020}. TEXES can observe at wavelengths ranging from 4.5 to 25\,$\mu$m, with a spectral resolving power ranging from 4000 to 80000 depending on the operating mode. The observations of Jupiter were centered around 8.02\,$\mu$m with a bandwidth of about 0.06\,$\mu$m where several spectral lines of the \ce{CH4} ($\nu_{4}$ band) P-branch lie \citep{brown_methane_2003}. During the March 2017 observing campaign, TEXES was used in its highest spectral resolution mode (R=80000).

The Gemini/TEXES observations were used to retrieve vertically resolved latitude/longitude temperature maps of Jupiter's stratosphere. These maps were compared to the lower spatial resolution maps from the IRTF to show that IRTF/TEXES is capable of fully resolving the meridional structure of the JEO \citep{cosentino_effects_2020}. The pressure range probed by the TEXES data ranges from 0.1 to 30\,mbar with a vertical resolution of approximately one scale height. Beyond this pressure range, the temperature vertical profiles converge toward the profile from \cite{moses_photochemistry_2005}, which is taken as a priori for the retrievals. The horizontal resolution is 2$^{\circ}$ in latitude and 4$^{\circ}$ longitude. The uncertainty on the retrieved temperatures is about 2\,K. 

\begin{figure}[h]
	\centering
	\includegraphics[width=9cm, keepaspectratio]{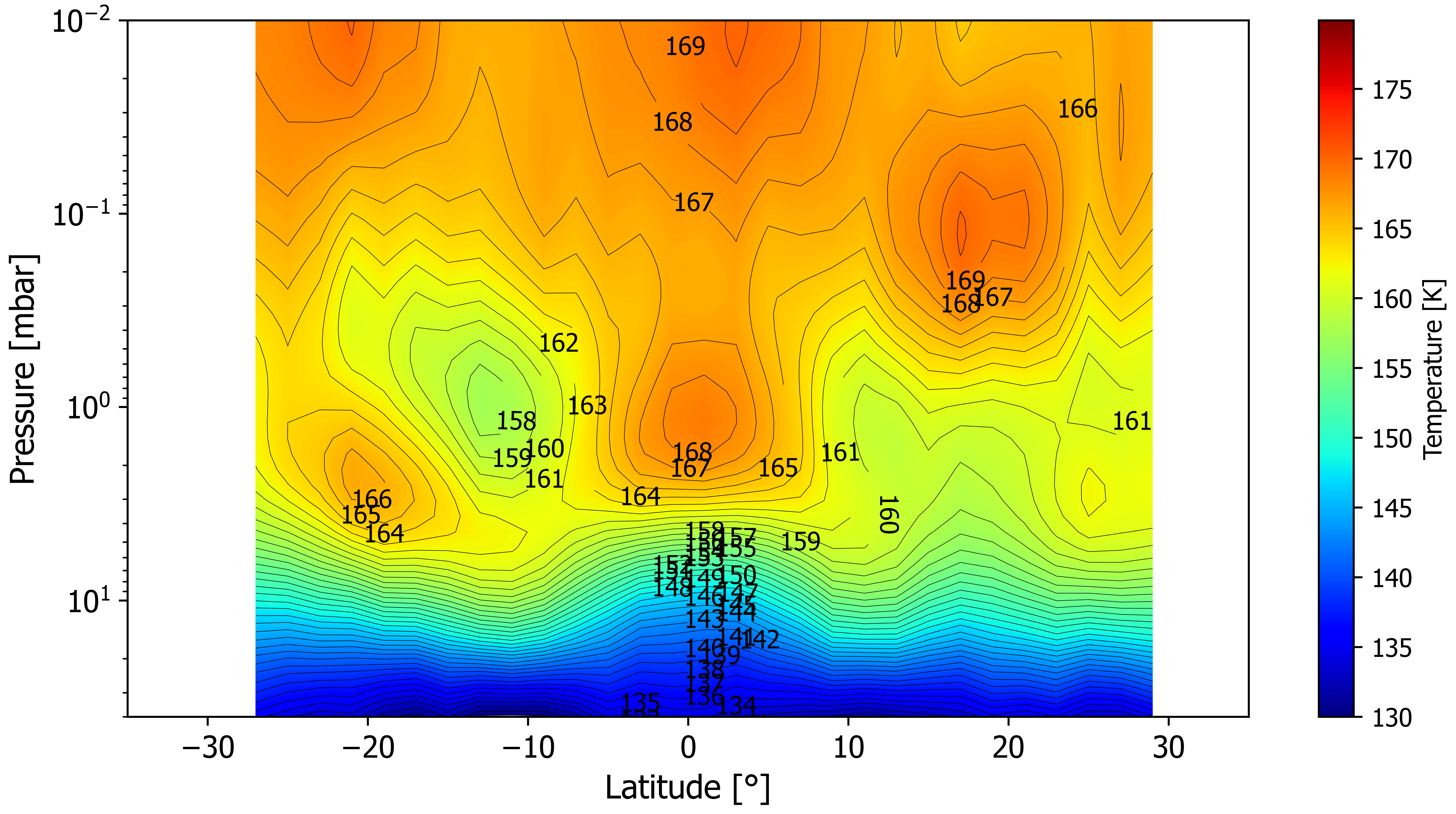}
	\caption{Average of the temperature fields at the eastern and western limbs covered by the ALMA wind observations. This temperature field is referred to as east-west limb mean in the paper and results from the average of the fields shown in Figures \ref{fig:temperature_east} and \ref{fig:temperature_west}. Latitudes are planetocentric.
	}
	\label{fig:temperature_mean_east-west}
\end{figure}

The Gemini/TEXES temperature field we use in this paper was retrieved from the combined data taken on 14, 16, and 20 March 2017, i.e., only 2-8 days apart from the ALMA wind measurements. By extracting the temperatures at the longitudes of the limbs probed by ALMA and accounting for the 15$^{\circ}$ longitudinal smearing, we produced an altitude-latitude temperature fields for each limb, and an average of both. The latter, referred to as the east-west limb mean in what follows, is shown in Figure \ref{fig:temperature_mean_east-west}. The former are shown in Figures \ref{fig:temperature_east} and \ref{fig:temperature_west}, and the zonal mean temperature field is shown in Figure \ref{fig:zonal_mean_temperature} of Appendix \ref{appendix:temps} for comparison. We note, however, that we only have a full longitudinal coverage over the eastern limb. Only one third of the western limb (155$^{\circ}$-170$^{\circ}$W) is covered by temperature measurements (155$^{\circ}$-160$^{\circ}$W).

\section{Models}
\label{section:models}

\subsection{Thermal wind equation}

Atmospheric dynamics can be interpreted with the equations of fluid mechanics. The so-called thermal wind equation (TWE) derives from Euler's equations for a frictionless fluid (e.g., \citealt{Pedlosky1979}). In spherical coordinates and assuming geostrophic equilibrium, the TWE relates the temperature gradient with the perpendicular velocity in the ({$\overrightarrow{e_{\theta}}; \overrightarrow{e_{\phi}}$}) plane\footnote{The unit vector $\overrightarrow{e_{\phi}}$ is oriented in the direction of the planet rotation such that positive winds are eastward.} of a fluid (such as an atmosphere) in a rotating frame. This equation is given by the following expression:
\begin{equation}
	\centering
		f_{0} \sin(\theta) \frac{\partial\overrightarrow{v}_{\perp}(r,\theta,\phi)}{\partial r} = \frac{r}{Tr_{0}}\overrightarrow{r}\wedge \overrightarrow{\bigtriangledown}_{\perp}T(r,\theta,\phi)
\end{equation}
where $f_{0}$ is the Coriolis parameter at the north pole, $\theta$ is the latitude varying from -90$^{\circ}$ to 90$^{\circ}$, $\overrightarrow{v_{\perp}}(r,\theta,\phi)$ is the horizontal fluid velocity at the planet surface, $r_{0}$ is the mean radius of the planet, and $T$ is the temperature field. From this equation, and after projecting on the zonal axis, we can easily relate the zonal wind speed with the latitudinal temperature gradient. Thus, we have the following expression: 
\begin{equation}
	\centering
 		\frac{\partial v_{\phi}}{\partial \ln(P)} = \frac{1}{ \sin(\theta)}\frac{R(P)}{f_{0}r_{0}}\frac{\partial T}{\partial \theta}
\end{equation}
where $P$ is the pressure and $v_{\phi}$ is the zonal wind velocity. $R(P)=\frac{k_{\text{B}}}{M(P)}$   is the specific gas constant of the Jovian atmosphere calculated for each altitude. $M(P)$ is the mean molecular mass of Jupiter's atmosphere as a function of pressure. We derive it from the model used in \cite{benmahi_monitoring_2020}. Establishing this equation assumes hydrostatic and geostrophic equilibrium, and the latter is guaranteed by the small Rossby number ($Ro$) in Jupiter's atmosphere. Since the Coriolis force vanishes at the equator because of the $f_{0}\sin(\theta)$ factor, this equation diverge at the equator $(\theta = 0)$. This is why \cite{flasar_intense_2004} used the TWE only down to latitudes of about 5$^{\circ}$ in their zonal wind derivation.

\subsection{Equatorial thermal wind equation}

\cite{marcus_equatorial_2019} derived an equatorial thermal wind equation (EQTWE). It uses the Laplacian in latitude of the temperatures, which enables getting rid of the $\frac{1}{\sin(\theta)}$ factor of the TWE. As a result, the EQTWE does not diverge at low latitudes and is thus particularly useful to replace the TWE in the equatorial zone. Its expression is given by:

\begin{equation}
 \frac{\partial v_{\phi}^{M}}{\partial \ln(P)} = \frac{R(P)}{f_{0}r_{0}}\frac{\partial^{2} T^{M}}{\partial \theta^{2}}
\end{equation}
where $T^{M}(P,\theta)=\frac{T(P,\theta)+T(P,-\theta)}{2}$ is the mirror-symmetric component of the temperature about the equator. The anti-mirror-symmetric component of the temperature is $T^{A}(P,\theta)=\frac{T(P,\theta)-T(P,-\theta)}{2}$. According to \cite{marcus_equatorial_2019}, the EQTWE has a fractional error $\propto \lvert \frac{T^{A}}{T^{M}} \rvert^{2}$ and it is valid in the latitudinal range +/-18$^{\circ}$ with an error of less than 10\%.

The derivation of the EQTWE requires the same assumptions as for the TWE, except the limitations regarding the Rossby number at the equator, and assumes that the flow is symmetrical around the equator (see Appendix A in \citealt{marcus_equatorial_2019}). 

\subsection{Assumptions and equation solving}

To carry out our study, we must also assume that Jupiter's temperature field remains stationary over the time interval between the TEXES thermal and ALMA wind measurements (from 2 to 8 days). This is justified for several reasons: The characteristic time of variability of cloud and storm dynamics in the troposphere is about a few days. However, their effects on stratospheric temperatures are transported by wave and energy propagation on time scales comparable to the periodicity of the JEO. Moreover, seasonal effects on Jupiter are weak (e.g., \citealt{hue_photochemistry_2018}) and the considered duration is negligible compared to Jupiter's year, and mostly the considered duration is much shorter than the radiative timescales in Jupiter's stratosphere \citep{guerlet_radiative-equilibrium_2020}.

\begin{figure}[h]
	\centering
	\includegraphics[width=10cm, keepaspectratio]{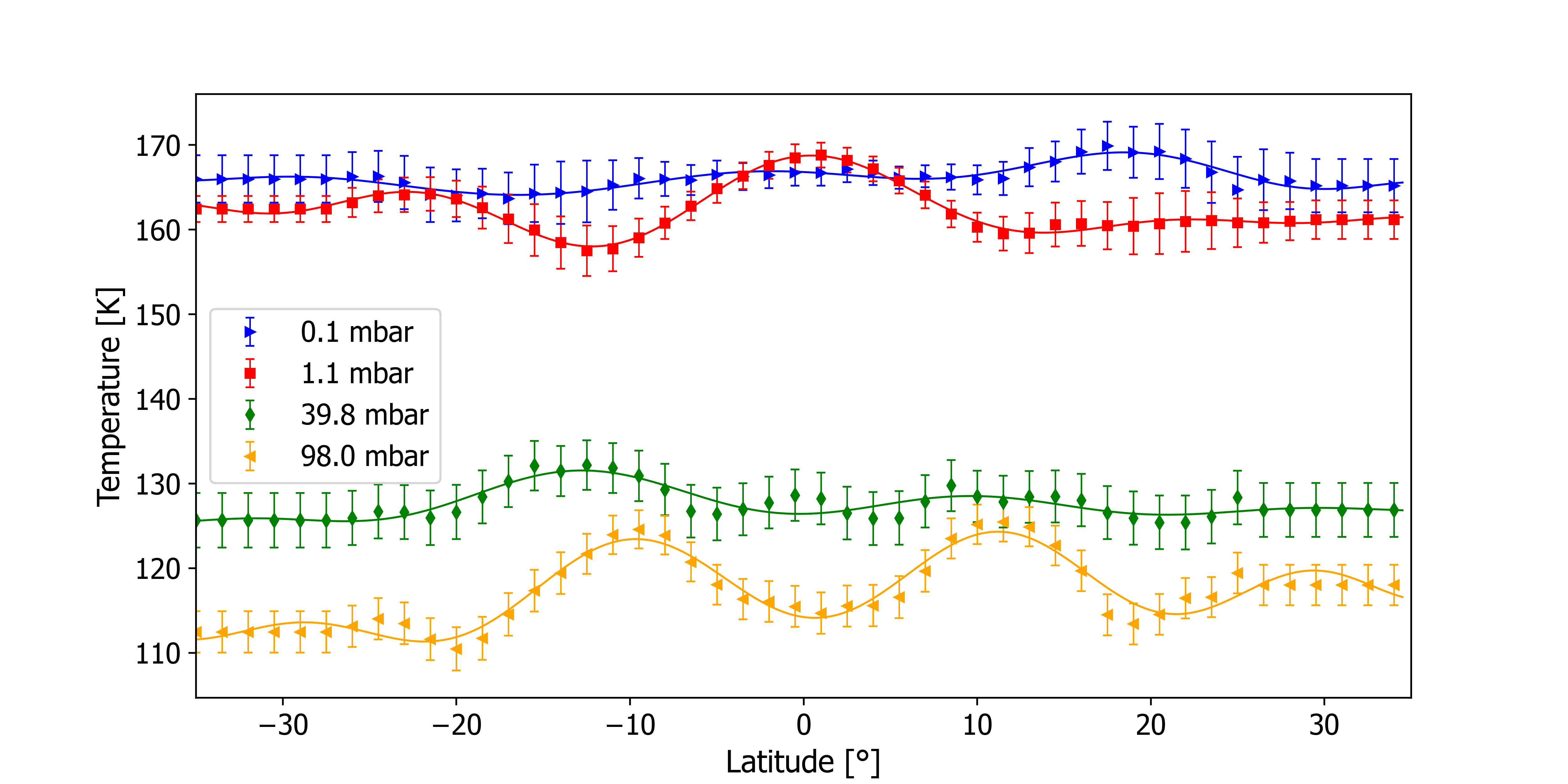}
	\caption{Examples of Legendre polynomial series fitting of the temperatures as a function of latitude for four different pressure levels. The two lower pressure profiles (0.1 and 1.1\,mbar) are from the \citet{giles_vertically-resolved_2020} dataset that we primarily use in this paper. The higher pressure profiles (39.8 and 98\,mbar) are from the \citet{fletcher_jupiters_2020} dataset. Fits are in solid lines and data are shown with symbols and error bars of the corresponding color.
	}
	\label{fig:temperature_fit_example}
\end{figure}

Before solving the TWE and EQTWE, we smoothed over latitude the temperature field and the 1-mbar wind speeds to obtain smooth and continuous derivatives (see Appendix \ref{appendix:smoothing}). We smoothed the various temperature fields (the east-west limb mean of Figure \ref{fig:temperature_mean_east-west}, the zonal average of Figure \ref{fig:zonal_mean_temperature}, the eastern and western limbs of Figures \ref{fig:temperature_east} and \ref{fig:temperature_west}) with a Legendre polynomial series up to degree 17. An example of fits at several pressures is shown in Figure \ref{fig:temperature_fit_example}. For the ALMA wind velocities at 1\,mbar, we used a Legendre polynomial series up to degree 35 (see Figure \ref{fig:wind_exemple_fit_legendre}). We determined the highest degree of the fitting polynomials such that the fits were within observation uncertainties. For temperatures and velocities, we used a latitudinal sampling of 0.25$^{\circ}$ and we solved the TWE and EQTWE with this sampling from 35$^{\circ}$S to 35$^{\circ}$N. We integrated the equations upwards and downwards starting with the ALMA wind velocities as initial conditions at $P_{0}=1$\,mbar.

Finally, we need to determine the latitude range around the equator in which we solve the EQTWE and then switch to the TWE. The TWE is highly dependent on $Ro$ and its fractional error is about $\propto Ro$. We thus estimate $Ro$ as a function of latitude by considering a characteristic velocity scale for Jupiter's atmosphere of 100\,m/s. We find $Ro$ $\sim$ 0.2 at +/- 5$^{\circ}$ latitude. So, we solve the EQTWE on the mirror-symmetric component of the temperature field $T^{M}(P,\theta)$ in the latitude interval [-3$^{\circ}$; 3$^{\circ}$], where the initial velocity condition $v_{\phi}(\theta,P_0 )$  (i.e., the fitted curve in figure \ref{fig:wind_fited}) is actually quasi-symmetrical about the equator. We then solve the TWE in the latitude interval [-35$^{\circ}$; -5$^{\circ}$] $\cup$ [5$^{\circ}$; 35$^{\circ}$] using $T(P,\theta)$. Finally, we use a bilinear interpolation between the two results in the [-5$^{\circ}$; -3$^{\circ}$] $\cup$ [3$^{\circ}$; 5$^{\circ}$] range, to combine the results in a single map.


\section{Results and discussion}
\label{section:results}

\subsection{Zonal winds in the JEO region}
\begin{figure}[h]
	\centering
	\includegraphics[width=9cm, keepaspectratio]{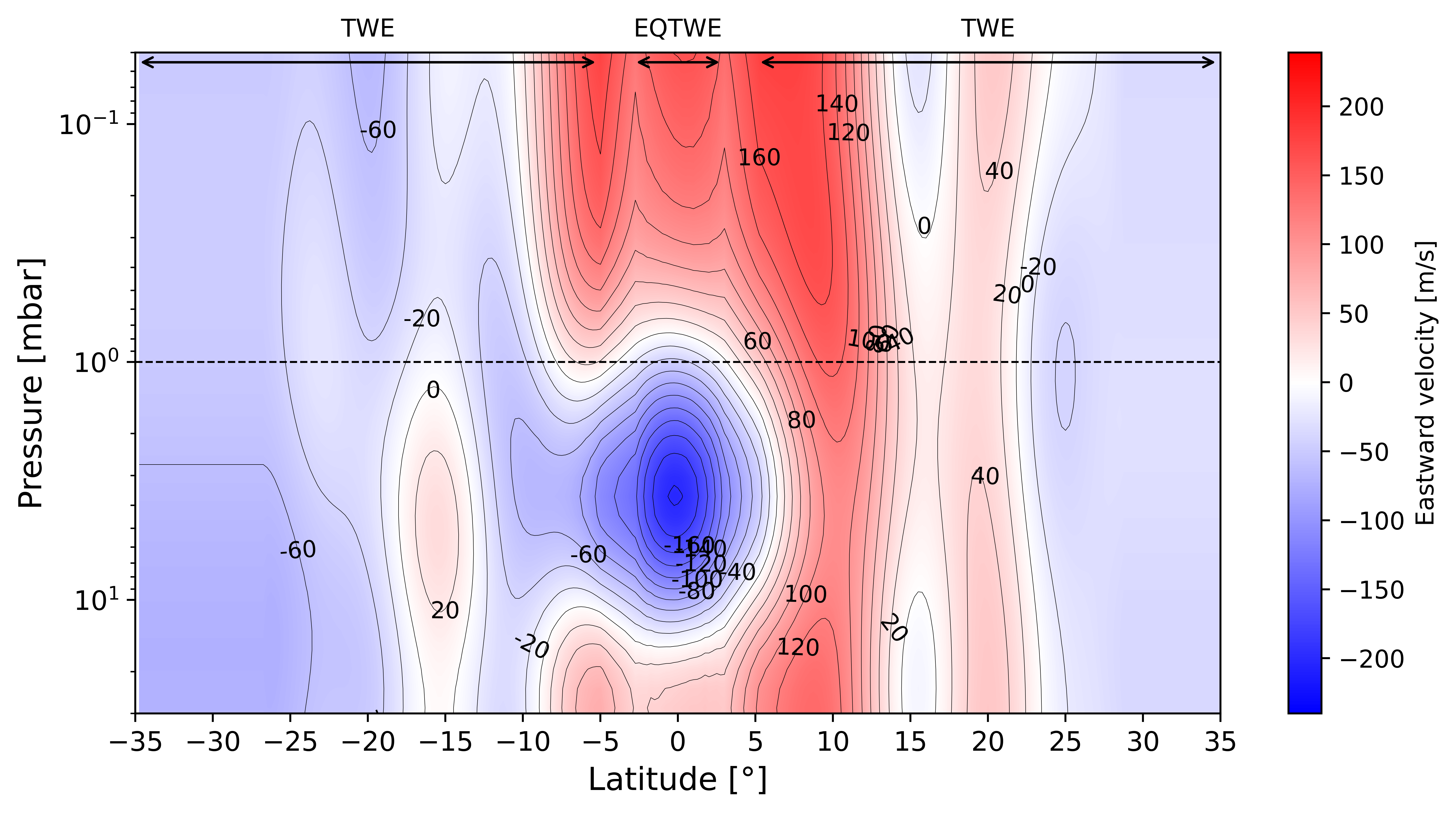}
	\includegraphics[width=9cm, keepaspectratio]{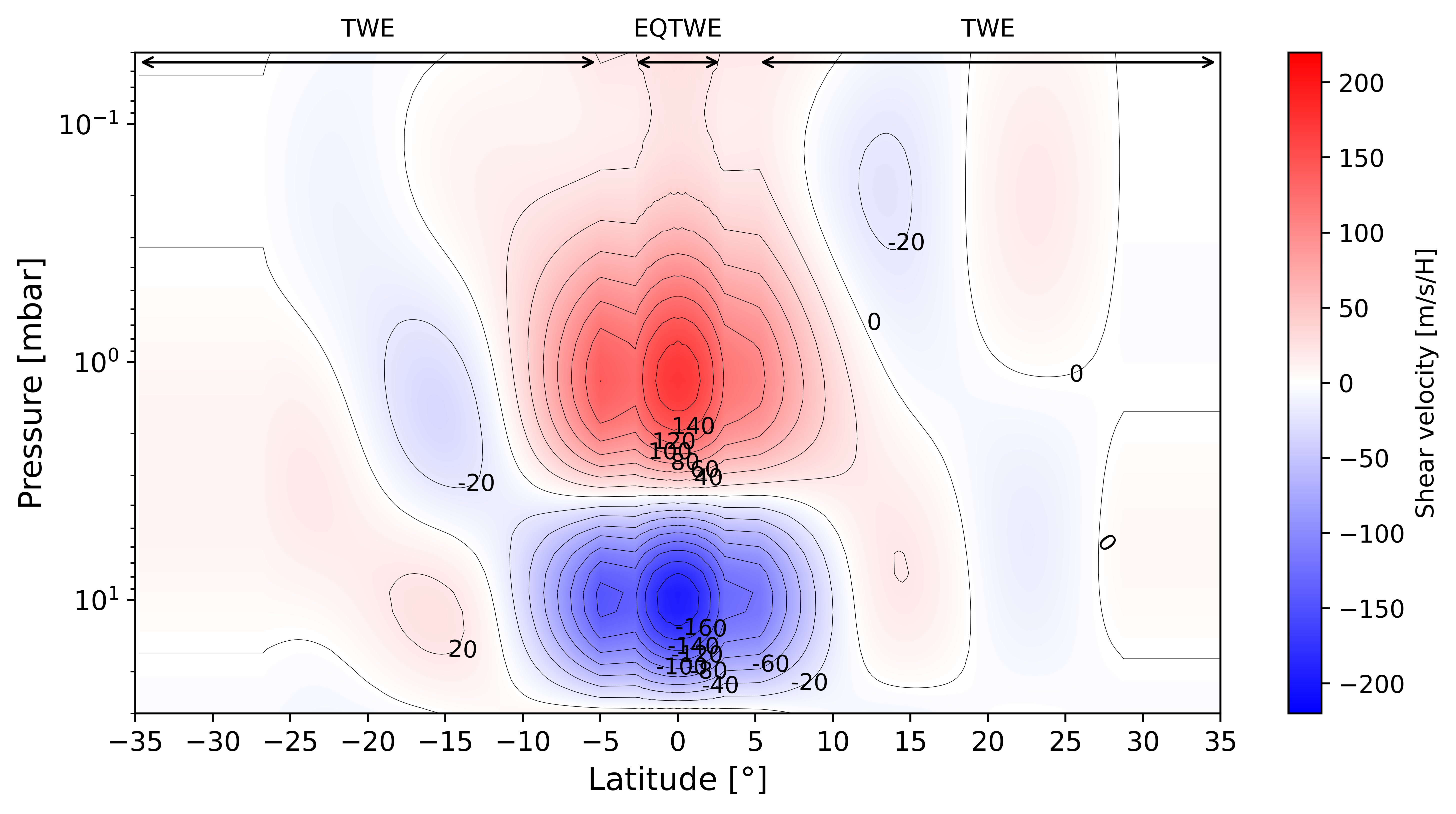}
	\caption{(Top) Eastward wind velocities derived from the east-west limb mean temperature map of Figure \ref{fig:wind_fited} and the measured winds at 1\,mbar of Figure \ref{fig:temperature_mean_east-west} with the EQTWE and TWE. The dashed horizontal line represents the altitude where stratospheric winds were measured by \cite{cavalie_first_2021}. (Bottom) Wind shear as obtained from the east-west limb mean temperature field.
	}
	\label{fig:wind_mean_limb}
\end{figure}

    The eastward wind velocities we derive from the east-west limb mean temperature map of Figure \ref{fig:temperature_mean_east-west} with the EQTWE and TWE and using the wind speeds measured with ALMA at 1\,mbar of Figure \ref{fig:wind_fited} as an initial condition, are shown in Figure \ref{fig:wind_mean_limb} (top). We note that there is no sharp discontinuity between the two equation solutions in the [-5$^{\circ}$; -3$^{\circ}$] $\cup$ [+3$^{\circ}$; +5$^{\circ}$] latitude range and from 0.5 to 30\,mbar. Above the 0.5\,mbar pressure level in the united latitude range [-5$^{\circ}$; -3$^{\circ}$] $\cup$ [3$^{\circ}$; 5$^{\circ}$], we notice small differences between the two equation solutions. Because the TWE is very dependent on the Rossby number, its validity also depends on it. Thus, the discrepancy between the TWE and the EQTWE is due to the local variability of the Rossby number.

We present these computations in the 0.05-30\,mbar range (Figure \ref{fig:wind_mean_limb} for example), where the TEXES observations are sensitive to temperatures and have the lowest uncertainties in the retrievals. The JEO can clearly be identified in the zonal wind map by vertically alternating zonal jets. We find a strong westward (i.e. retrograde) jet at the equator centered at about 4\,mbar with a peak velocity of 200\,m/s. This jet has a full-width at half-maximum of about 7$^{\circ}$ in latitude and 50\,km in altitude (two scale heights). At pressures lower than 1\,mbar, we find an eastward jet, which is 20$^{\circ}$ wide in latitude and having a full-width at half-maximum of about 80\,km in altitude (between 0.05 and 0.5\,mbar). The vertical stratification about the equator is also unambiguously characterized, with a peak-to-peak difference of 300\,m/s between 4\,mbar and 0.1\,mbar. The JEO jet in March 2017 is almost perfectly in opposition of phase compared to the state observed by \cite{flasar_intense_2004} in December 2000 at the time of the Cassini flyby. The amplitude of the JEO jet in these two observations are comparable (200\,m/s vs. 140\,m/s), even though the exact velocity at the equator is not known in \cite{flasar_intense_2004} because of the limitations of the TWE at the equator. At $\sim$10$^{\circ}$N, the eastward jet is vertically extended over the entire pressure range with an average velocity of 125\,m/s. Beyond +/-15$^{\circ}$, winds have amplitudes lower than 40\,m/s.

In Figure \ref{fig:wind_mean_limb} (bottom), we map the wind shear ($\frac{\partial v_{\phi}}{\partial r}$), as obtained from the east-west limb mean temperature field (Figure \ref{fig:temperature_mean_east-west}). We can clearly see two wind shear spots, centered around the equator, with positive and negative amplitudes of $\sim$160\,m/s/H respectively above and below the $\sim$4\,mbar pressure level where the westward jet is located (Figure \ref{fig:wind_mean_limb}). The two wind shear spots have a full-width at half-maximum of about 12$^{\circ}$ in latitude. Beyond +/-10$^{\circ}$ of latitude, the wind shear is negligible. Such vertical and latitudinal extensions, as well as amplitudes, are comparable to previous estimates \citep{fletcher_mid-infrared_2016,marcus_equatorial_2019}.

The zonal wind map we obtained from the Gemini/TEXES measurements in March 2017 combined to the 1\,mbar zonal wind measured by \cite{cavalie_first_2021} using ALMA and the result obtained by \cite{fletcher_mid-infrared_2016} from the zonal temperature field measured by IRTF/TEXES in December 2014, are almost in opposition of phase in the 1 to 10\,mbar range.
The time interval between the two measurements is $\Delta T$ $\sim$ 2 years and 4 months. This would lead to a JEO period of ~4 years and 8 months, in agreement with previous measurements \citep{leovy_quasiquadrennial_1991,orton_thermal_1991}. We note that \cite{antunano_long-term_2020} has now demonstrated that this periodicity is variable and can even be disrupted, as observed by \cite{giles_vertically-resolved_2020}. Such disruptions may originate from the outbreak of thermal anomalies, like the one seen in May 2017 at 20$^{\circ}$N, 1\,mbar at 180$^{\circ}$W (See Fig. 7 in \citealt{giles_vertically-resolved_2020}). 

\subsection{Longitudinal variability of zonal winds in the JEO region}

\begin{figure}[h]
	\centering
		\includegraphics[width=9cm, keepaspectratio]{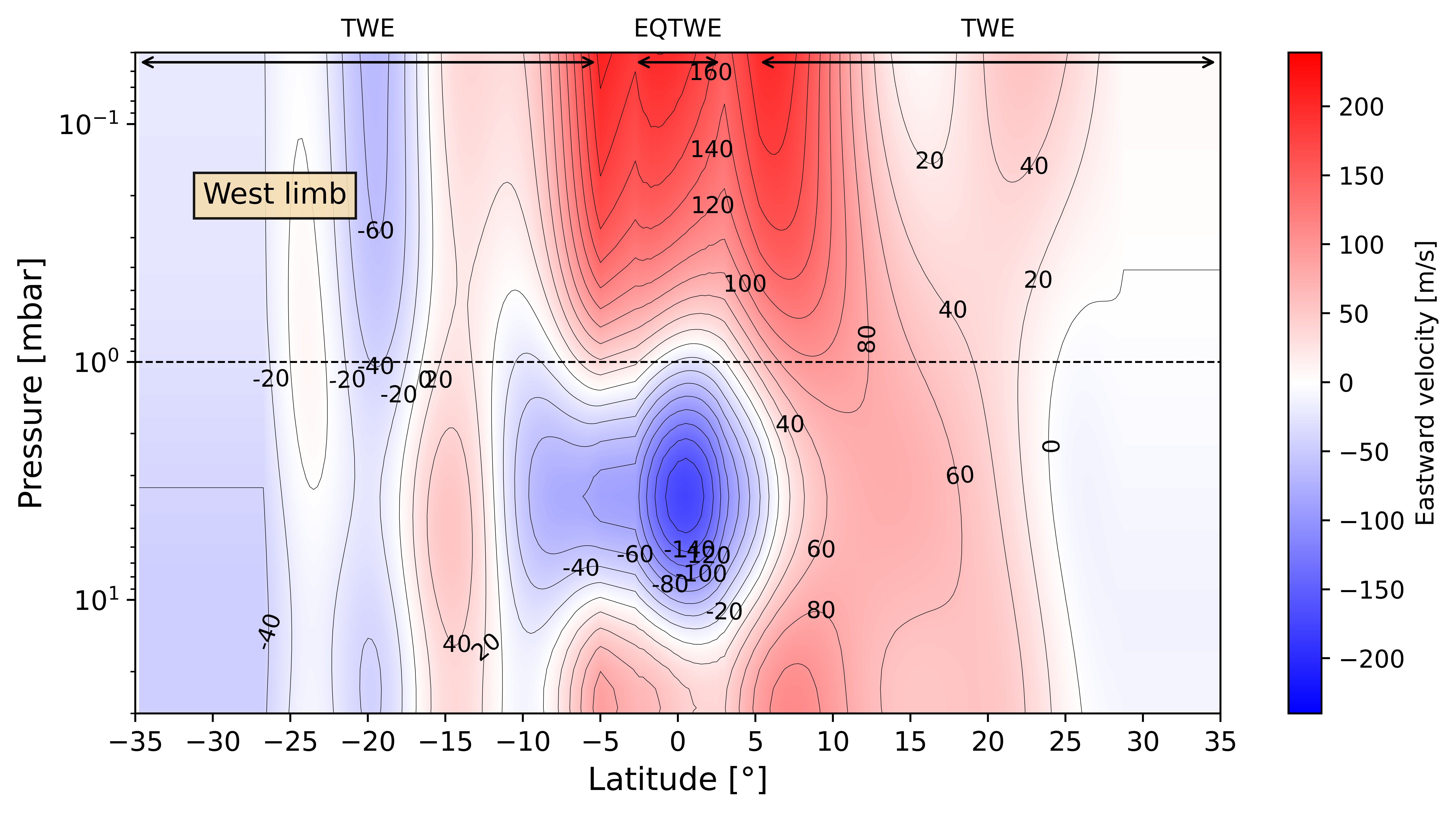}
		\includegraphics[width=9cm, keepaspectratio]{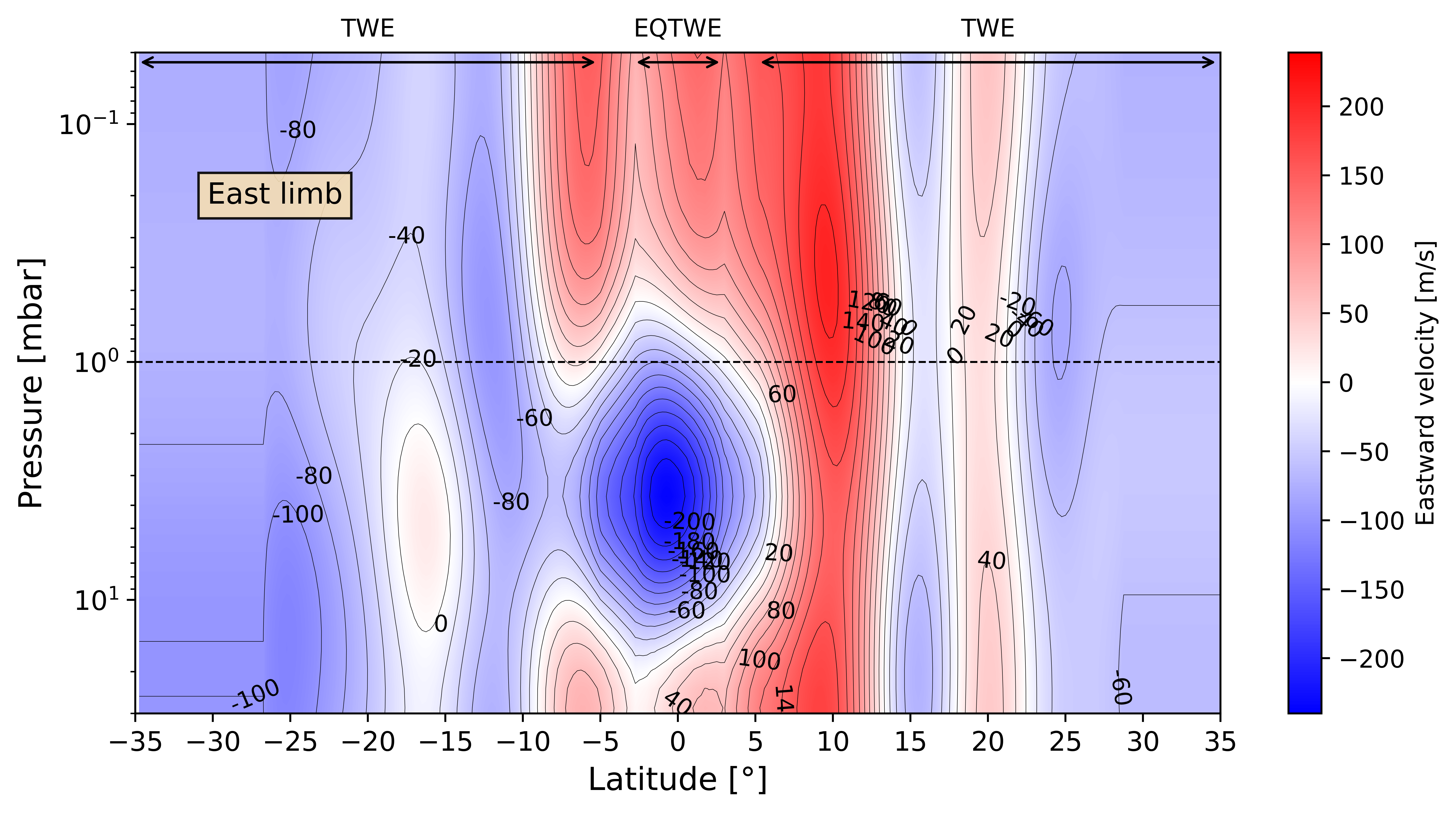}
	\caption{Eastward zonal wind velocities mapped independently at the western (top) and the eastern (bottom) limbs.}
	\label{fig:wind_map_east_west}
\end{figure}

The wind velocity map we derive from the temperature zonal mean, from the eastern limb only, and from the western limb only (Figures \ref{fig:zonal_mean_temperature}, \ref{fig:temperature_east}, and \ref{fig:temperature_west}, respectively), are presented in Figures \ref{fig:zonal_mean_wind_map}, and \ref{fig:wind_map_east_west}, respectively. The two latter wind maps are obtained by using the eastern and western ALMA wind measurements of Figure \ref{fig:wind_east-west} as initial condition. 

We notice in Figure \ref{fig:wind_map_east_west} that the 10$^\circ$N wind peak observed at 1\,mbar with ALMA can be tracked down to the lower stratosphere, where it is centered around 7$^\circ$N. This latitude is the latitude of the northern peak of the double-horned structure observed around the equator at the cloud-top \citep{barrado-izagirre_jupiters_2013}. The eastward jet at 10$^{\circ}$N at 1\,mbar and the one at 7$^{\circ}$N and 30\,mbar seem to be linked. Indeed, the eastward column connecting the two altitudes (1 and 30\,mbar) seems to be distorted by the central westward jet at 4\,mbar. We think that these two peaks are correlated and connected vertically, and that the presence of the westward jet, and thus of the planetary wave generating the JEO, results in a latitudinal shift between the two eastward peaks at 1\,mbar and at 30\,mbar. The peak at 7$^{\circ}$N and 30\,mbar is likely tied to the cloud top northern branch of the double-horned structure mentioned above.

A subject of debate in \citet{cavalie_first_2021} was the limb-to-limb velocity difference at 10$^\circ$N, which the authors tentatively attributed to local vortices. The lack of full coverage of the western limb temperatures prevents us from settling this claim, although the limited data we have seem to indicate the presence of a hot spot between 10$^{\circ}$N and 15$^{\circ}$N and between 1 and 2\,mbar as seen in Figure \ref{fig:west-zonal_temperature} of Appendix \ref{appendix:tempdiffs}. This hot spot, if anticyclonic, would decelerate the winds about $\sim$10$^{\circ}$N and accelerate the wind about $\sim$20$^{\circ}$N, in qualitative agreement with the ALMA data. Eastward-westward wind velocities at the northern and southern boundaries of an anticyclonic feature can reach 100-150\,m/s, as in the Great Red Spot \citep{choi2007}. In the stratosphere, the most famous example of anticyclonic feature was observed in Saturn's Great Storm in 2010-2011. At 2\,mbar, \citet{fletcher2012} estimated zonal wind velocities at the northern and southern boundaries of the vortex in the 200-400\,m/s range. So, the 100\,m/s lower velocity observed on the 10$^\circ$N jet on the western limb in the ALMA data (compared to the eastern limb) could be at least partly explained by the presence of an anticyclonic feature centered at 15$^\circ$N.

The winds obtained from the east-west limb mean (Figure \ref{fig:wind_mean_limb}) and those obtained from the zonal mean (Figure \ref{fig:zonal_mean_wind_map}) are very similar in the +/-10$^{\circ}$ latitude range, indicating that the east-west limb mean both in winds and temperatures at 1\,mbar was a fair representation of the zonal mean on this occasion. The westward equatorial jet at 4\,mbar has a similar shape and amplitude. Outside this range, the differences are less than 20\,m/s, which is about the HCN wind measurement uncertainty with ALMA. 

More significant differences arise when we compare the winds obtained from the two limb temperatures independently (Figure \ref{fig:wind_map_east_west}). The westward equatorial jet at 4\,mbar is 50\,m/s stronger on the eastern limb than on the western limb. On the eastern limb, we also notice that the northern equatorial branch centerd around 7$^{\circ}$-10$^{\circ}$N of the upper stratospheric eastward jet is 50-75\,m/s stronger than in the western limb. Both extend down to the lower boundary of our calculations (30\,mbar). The distinct eastward barotropic jet at 20$^{\circ}$N disappears on the western limb.  

By comparing the 4\,mbar equatorial westward jet velocities in the two limbs (Figure \ref{fig:wind_map_east_west}), we find a difference of about 50\,m/s. This results from a combination of the differences between the velocities measured at 1\,mbar and the temperatures in the two limbs. The differences we find in the equatorial temperatures between the two limbs and between 1 and 4\,mbar is twice the longitudinal standard deviation in this pressure range.


\subsection{Equatorial cloud-top wind structure}
\begin{figure}[h]
	\centering
	\includegraphics[width=9cm, keepaspectratio]{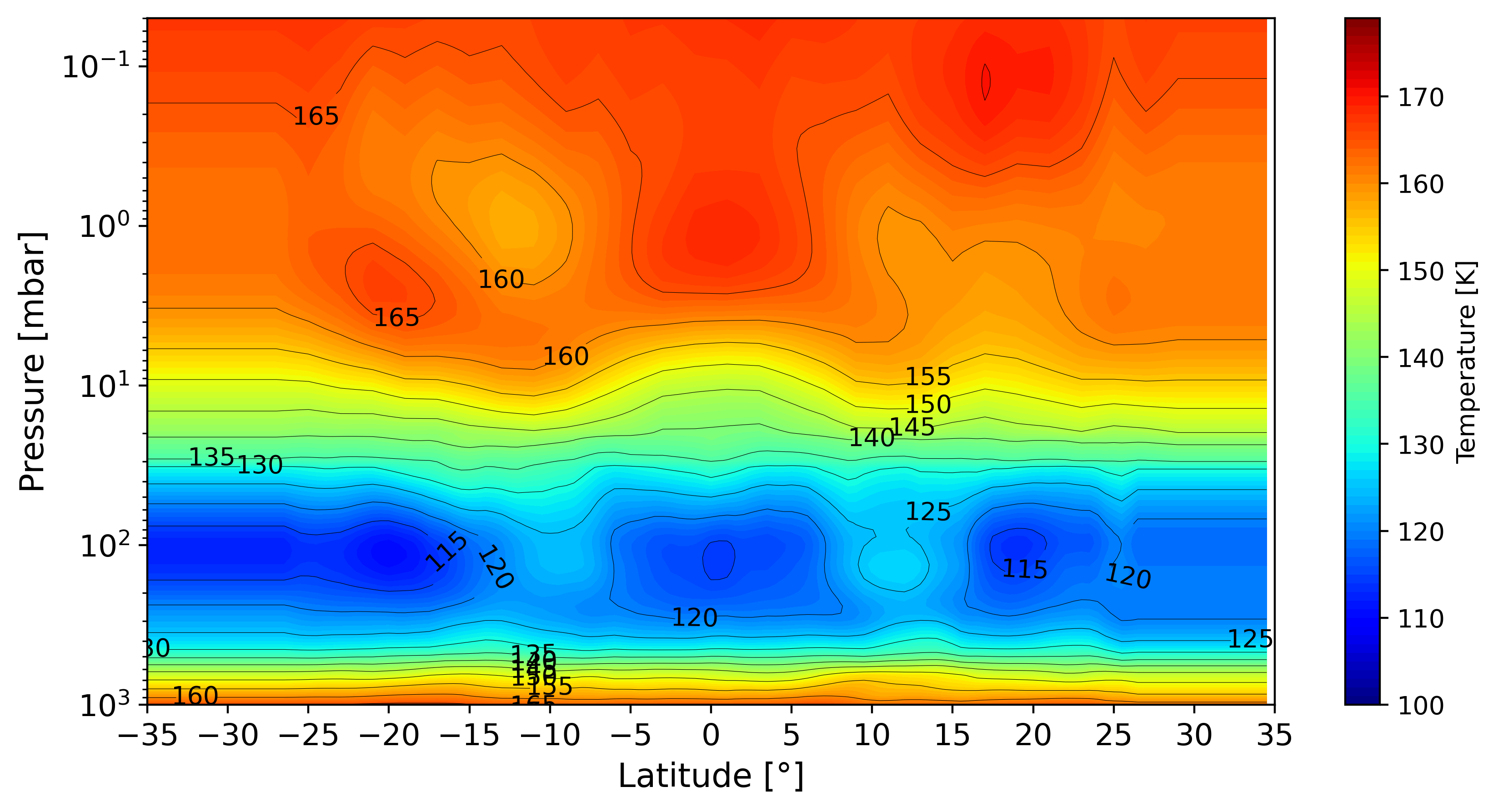}
	\caption{Temperature field resulting form the combination of retrievals obtained from the high spectral resolution observations of \citet{cosentino_effects_2020} (Figure \ref{fig:temperature_mean_east-west}) and the retrievals from lower spectral resolution observations of \cite{fletcher_jupiters_2020}, all performed between 12 and 20 March, 2017. This temperature field covers the 0.05-1000\,mbar pressure range.
	}
	\label{fig:temperature_complete_mean_limb}
\end{figure}

We checked whether the temperature and wind observations combined with our model allow the derivation of the cloud-top wind structure. We thus extended our east-west limb mean temperature map (Figure \ref{fig:temperature_mean_east-west}) down to upper tropospheric altitudes. To do so, we used the upper tropospheric and lower stratospheric temperatures as retrieved by \cite{fletcher_jupiters_2020} from lower spectral resolution Gemini/TEXES observations taken on 12-14 March 2017 and probed in the pressure range p<1000\,mbar. We combined the two temperature fields by averaging them between 20 and 30\,mbar for the relevant longitudes. This pressure range is chosen so as to minimize the overlap and thus favor the temperatures retrieved from the high spectral resolution observations at least down to the 20-mbar pressure level. The resulting thermal map is shown in Figure \ref{fig:temperature_complete_mean_limb} and covers pressures from 0.05 to 1000\,mbar. We then applied our thermal wind model, still using the ALMA wind measurements as the initial condition at 1\,mbar. 

\begin{figure}[h]
	\centering
	\includegraphics[width=9cm, keepaspectratio]{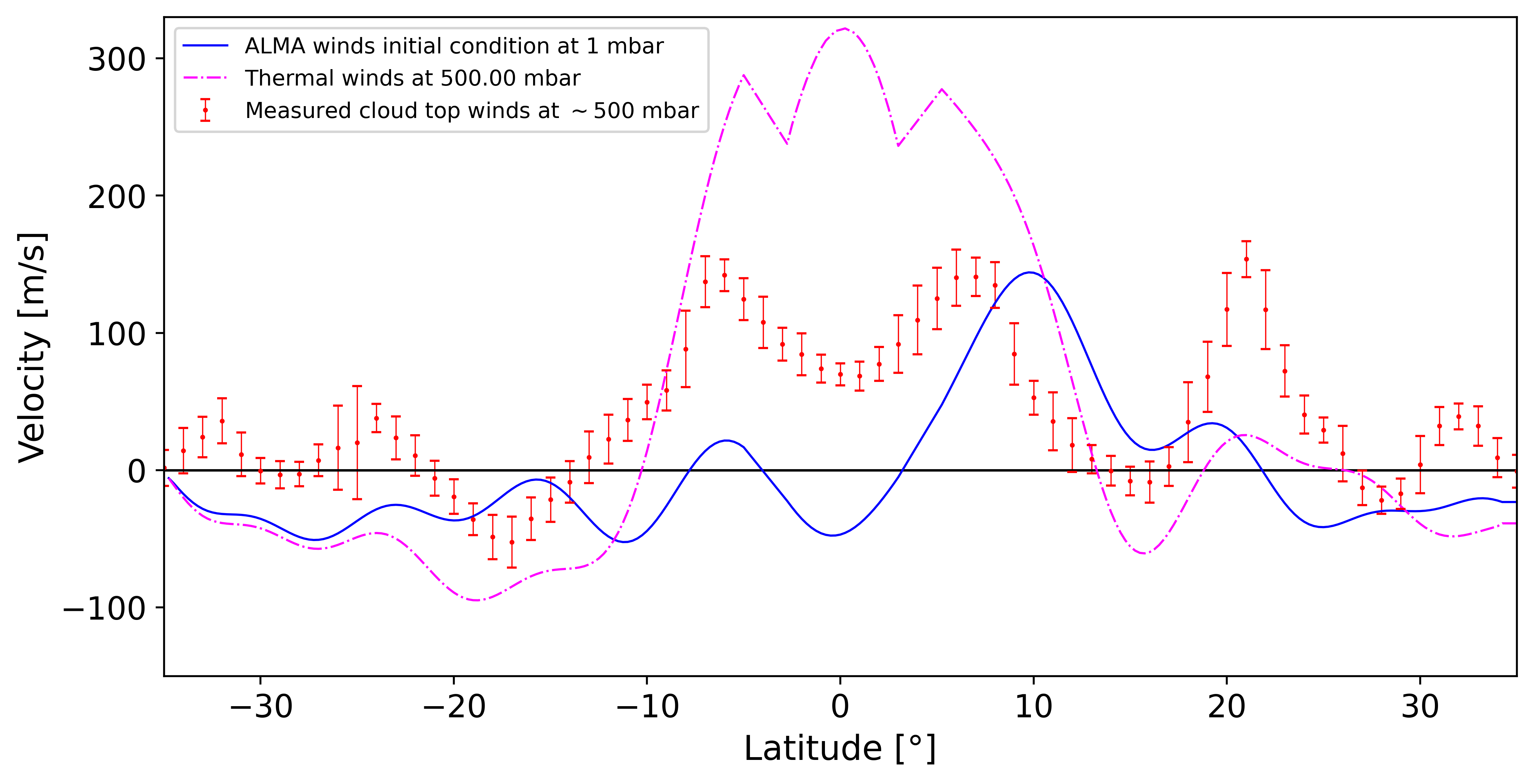}
	\caption{Comparison between the cloud-top winds speeds
measured at 500\,mbar in the visible (red points with error bars) with
the thermal winds (dashed magenta line) derived from the east-west
mean temperature field of Figure \ref{fig:temperature_mean_east-west} and the 1mbar
wind observations with ALMA of Figure \ref{fig:wind_fited} (blue solid line).
	}
	\label{fig:wind_extracted_troposphere}
\end{figure}

\begin{figure}[h]
	\centering
	\includegraphics[width=8cm, keepaspectratio]{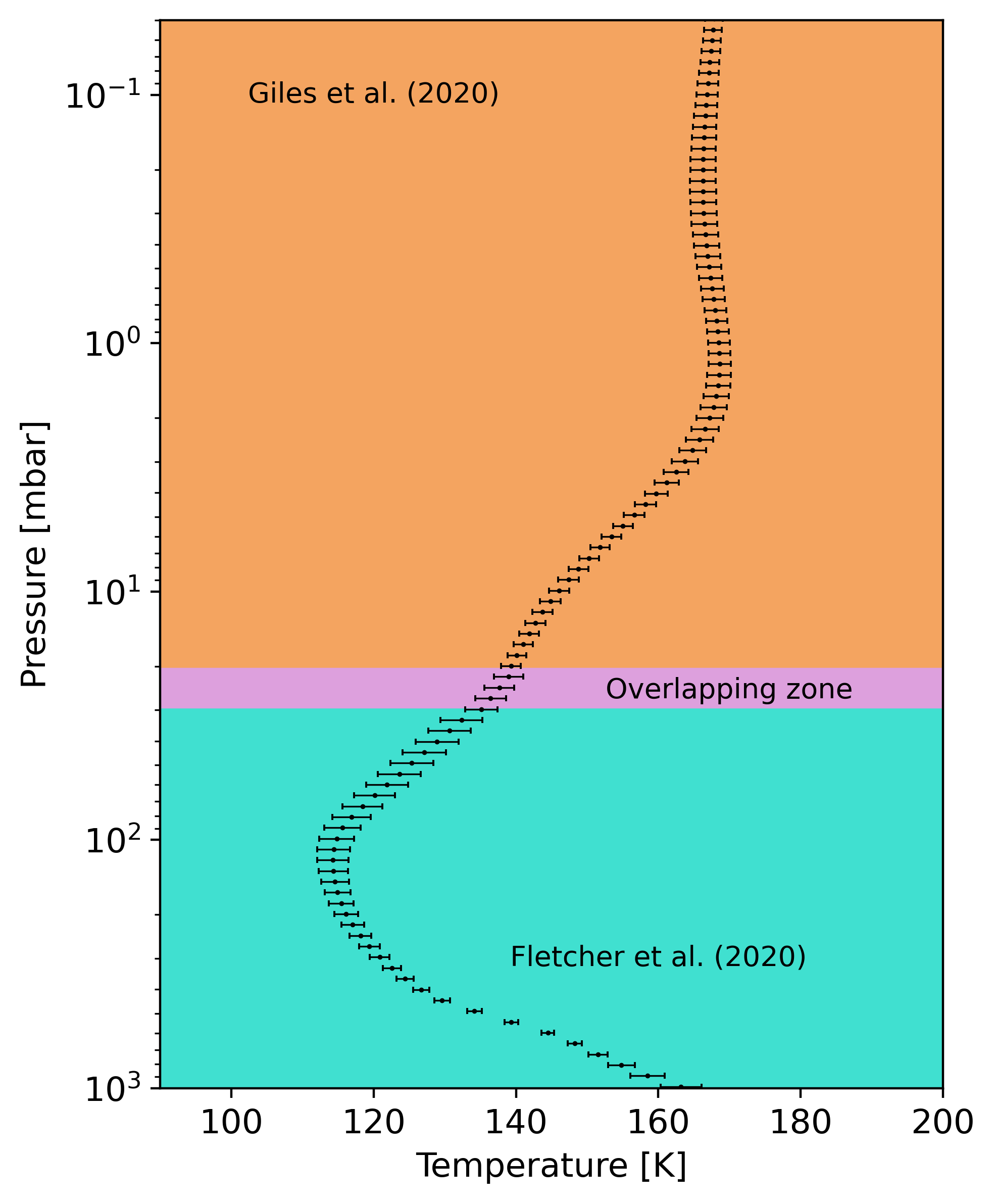}
	\caption{Vertical profile of the temperature at the equator extracted from the field of Figure \ref{fig:temperature_complete_mean_limb} from the combination of retrievals obtained from the high spectral resolution observations of \citet{giles_vertically-resolved_2020} (Figure \ref{fig:temperature_mean_east-west}) and the retrievals from lower spectral resolution observations of \cite{fletcher_jupiters_2020}, all performed between 12 and 20 March 2017. The two profiles are averaged in the 20-30\,mbar pressure range (refered to as ``overlapping zone'' on the plot).
	}
	\label{fig:temp_vert_prof}
\end{figure}

In Figure \ref{fig:wind_extracted_troposphere}, we compare the zonal wind profile calculated at 500\,mbar with the cloud-top wind observations performed in the visible range (e.g. \citealt{barrado-izagirre_jupiters_2013}). 
Our thermal wind results at 500\,mbar show a strong wind speed increase within $\pm$10$^\circ$, as expected from observations. However, we do not reproduce the double-horned shaped centered about the equator. Instead, we see that the TWE and EQTWE do not provide consistent results in the [-5$^{\circ}$; -3$^{\circ}$] $\cup$ [3$^{\circ}$; 5$^{\circ}$] ranges because of the longer vertical integration that causes larger deviations between the two solutions. In addition, the wind speeds around the equator are overestimated by a factor of $\sim$2 within $\pm$10$^\circ$. This deviation probably arises from the different vertical resolutions in the two temperature retrievals we combined to perform these computations. The lower vertical resolution of the upper tropospheric-lower stratospheric temperatures is, in turn, caused by the lower spectral resolution of the observations of \citet{fletcher_jupiters_2020} compared to those of \citet{giles_vertically-resolved_2020}. In addition, the deviation may also result from the higher uncertainties in the temperatures retrieved by \citet{fletcher_jupiters_2020} (on average $\pm$4\,K compared to the $\pm$2\,K of \citealt{giles_vertically-resolved_2020}). These higher uncertainties in the higher pressures can be better seen in Figure \ref{fig:temp_vert_prof}, wherer we present a vertical temperature profile at the equator resulting from the combination of the temperature fields of \citet{giles_vertically-resolved_2020} and \citet{fletcher_jupiters_2020}. These higher uncertainties in the higher pressures can also been seen in Figure \ref{fig:temperature_fit_example}. We finally tried to integrate the temperature field starting from the cloud-top wind speeds. The wind speeds we obtain at 1\,mbar (not shown here) are in total disagreement with the ALMA observations. 


\section{Conclusion} 
\label{section:conclusion}
The main outcomes of this paper can be summarized as follow:


\begin{itemize}

\item We have used the recent and first measurements of the jovian stratospheric winds obtained from ALMA observations \citep{cavalie_first_2021}, with the temperature field obtained nearly-simultaneously in March 2017 in the mid-infrared from Gemini/TEXES observations \citep{giles_vertically-resolved_2020}, to derive the zonal wind field as a function of pressure and latitude, in the equatorial zone of Jupiter's stratosphere where the Jupiter equatorial oscillation occurs.

\item We have used the thermal wind equation, complemented by the equatorial thermal wind equation of \citet{marcus_equatorial_2019} for the latitudes about the equator, to derive the Jovian stratospheric zonal winds from 0.05 to 30\,mbar and from 35$^\circ$S to 35$^\circ$N.

\item We derive the absolute stratospheric zonal wind speeds $\pm$35$^\circ$ about the equator, where the JEO takes place. We thus provide the general circulation modeling community with the first full diagnostic of the JEO zonal winds for a given date.

\item  In March 2017, we find a strong westward (i.e., retrograde) jet centered on the equator and about the 4\,mbar level with a peak velocity of 200\,m/s. The vertical stratification of the JEO winds is demonstrated and we find that the westward jet lies beneath a broader eastward (i.e., prograde) jet and the peak-to-peak contrast is $\sim$300\,m/s. 



\item We find longitudinal variability at the level of $\sim$50\,m/s when comparing the winds derived independently from the eastern and western limbs of the ALMA observations, even though the overall structure of the JEO remains similar.


\item When extending our zonal wind computations to the cloud-top by using complementary thermal data (also taken over the same time period), we tentatively find a global wind structure close to observations. We find a strong equator-centred prograde jet. However, the lower spectral resolution of the lower stratospheric and upper tropospheric temperature observations prevent us from a closer and more quantitative agreement. We neither recover the double-horned equatorial shape nor the 20$^\circ$N jet.


\end{itemize}

Such direct stratospheric wind and temperature measurements, performed nearly simultaneously open up a new and promising window to characterize and understand the Jupiter equatorial oscillation and, more globally, its general circulation. Repeated observations, on various timescales, are now needed to accomplish this promise. These can be achieved first with ALMA and ground-based infrared facilities, and later on with the Submillimetre Wave Instrument aboard the Jupiter Icy Moons Explorer. The technique presented in this paper can certainly be adapted to the other giant planets to study their general circulation and equatorial oscillations.

\begin{acknowledgements}
This work was supported by the Programme National de Plan\'etologie (PNP) of CNRS/INSU and by CNES. The authors thank L. N. Fletcher for providing them with the temperature retrievals of his 2016 paper. Coauthors Guerlet and Spiga acknowledge funding from Agence Nationale de la Recherche (ANR) project EMERGIANT ANR-17-CE31-0007.
\end{acknowledgements}

\bibliographystyle{aa} 
\bibliography{biblio.bib} 

\appendix

\section{\\Thermal wind velocities from alternative temperature maps \label{appendix:temps}}

\begin{figure}[ht]
	\centering
	\includegraphics[width=9cm, keepaspectratio]{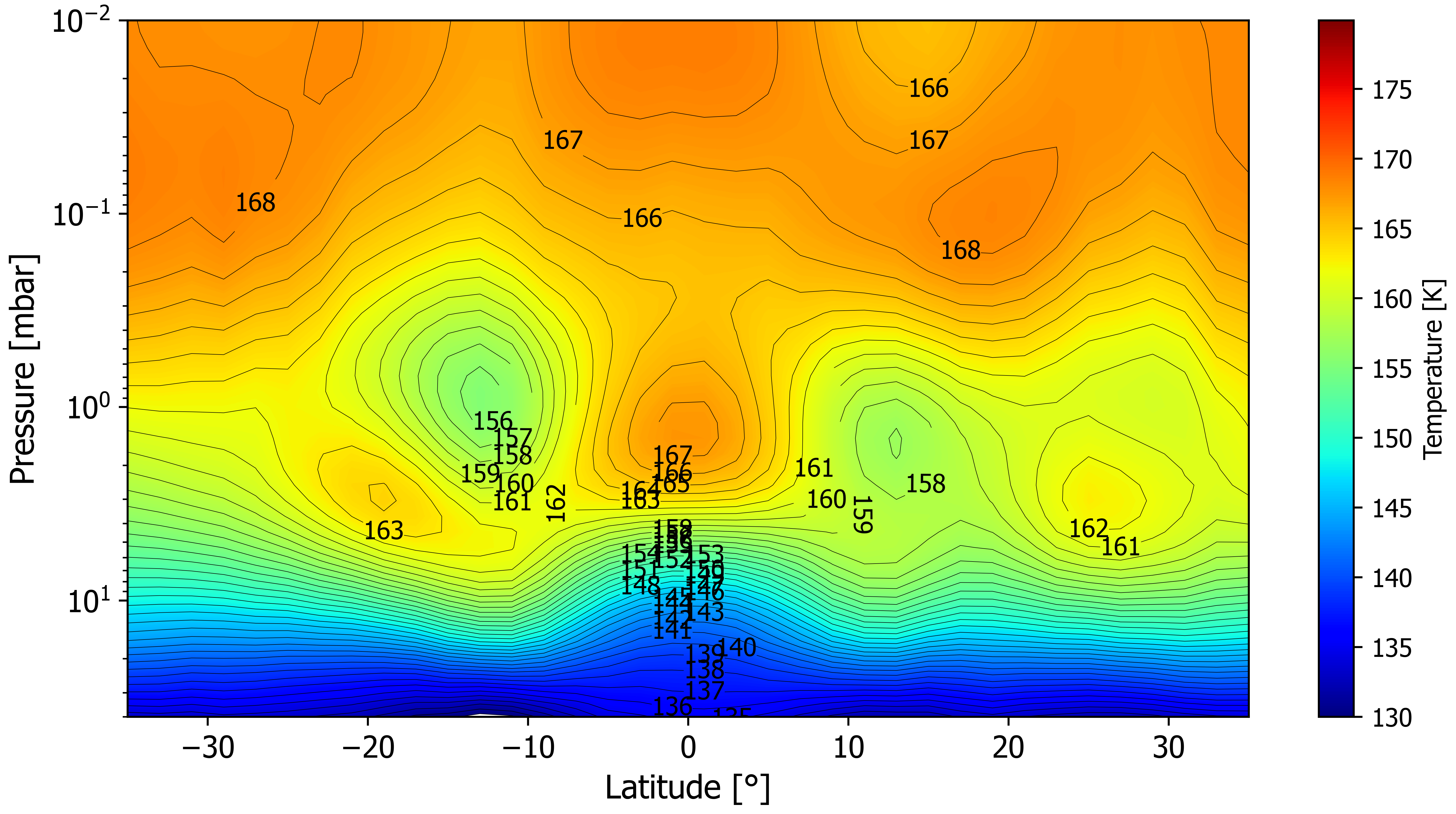}
	\caption{Zonal mean of the Jovian stratospheric temperature field, as observed with Gemini/TEXES on 14, 16, and 20 March, 2017.}
	\label{fig:zonal_mean_temperature}
\end{figure}

\begin{figure}[ht]
	\centering
	\includegraphics[width=9cm, keepaspectratio]{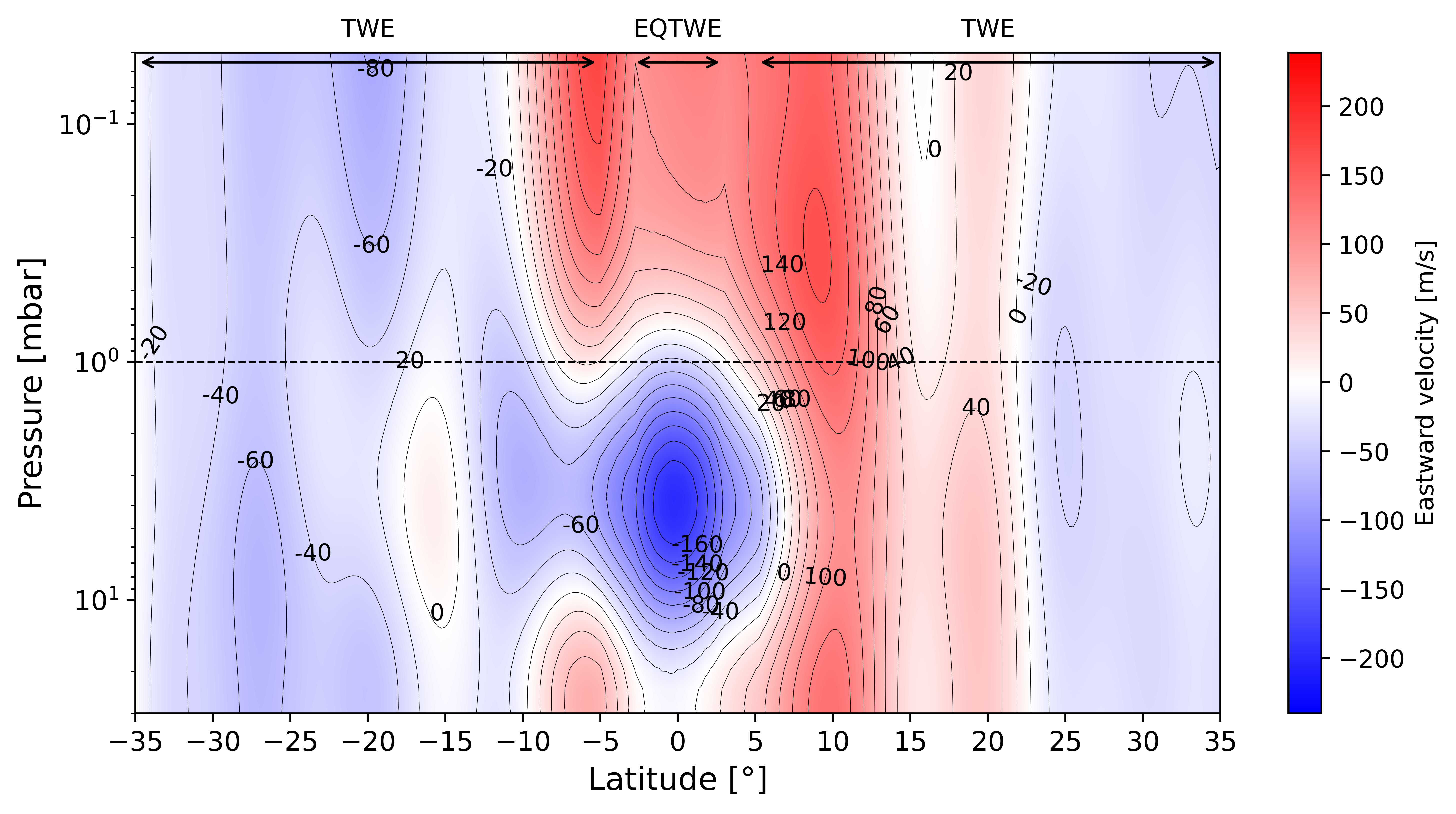}
	\caption{Eastward zonal wind velocities derived from the zonal mean temperature field of Figure \ref{fig:zonal_mean_temperature} and from the 1\,mbar winds of Figure \ref{fig:wind_fited}.}
	\label{fig:zonal_mean_wind_map}
\end{figure}

Figure \ref{fig:zonal_mean_temperature} presents the zonal mean of the temperatures in Jupiter's stratosphere from 0.01 to 30\,mbar on 12-20 March 2017. The corresponding wind velocity map is shown in Figure \ref{fig:zonal_mean_wind_map}. 

\begin{figure}[ht]
	\centering
		\includegraphics[width=9cm, keepaspectratio]{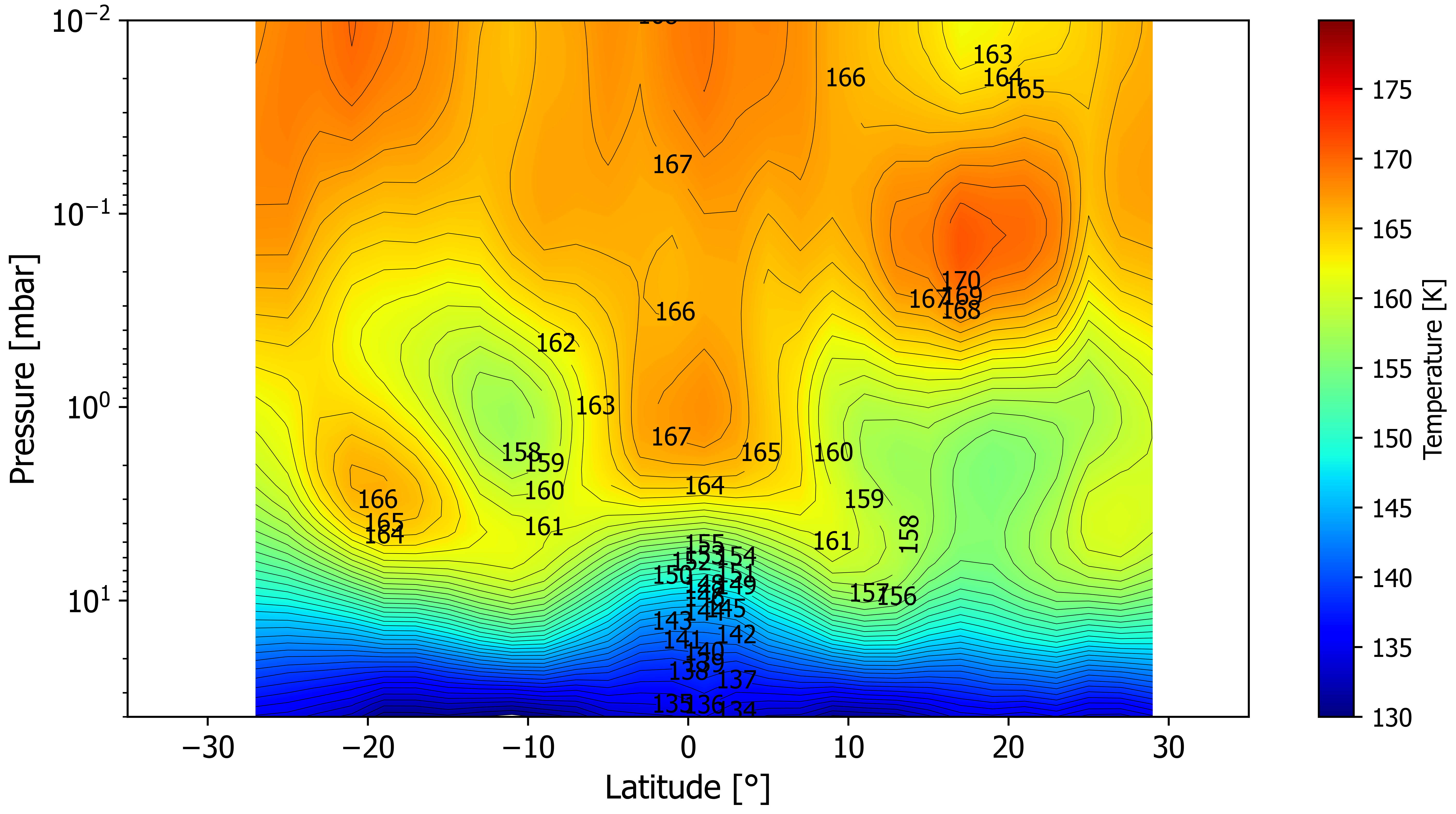}
		\caption{Eastern limb temperature field resulting from the average between 335$^{\circ}$ and 350$^{\circ}$ longitudes, as observed with Gemini/TEXES on 14, 16, and 20 March 2017.}
		\label{fig:temperature_east}
\end{figure}

\begin{figure}[ht]
	\centering
		\includegraphics[width=9cm, keepaspectratio]{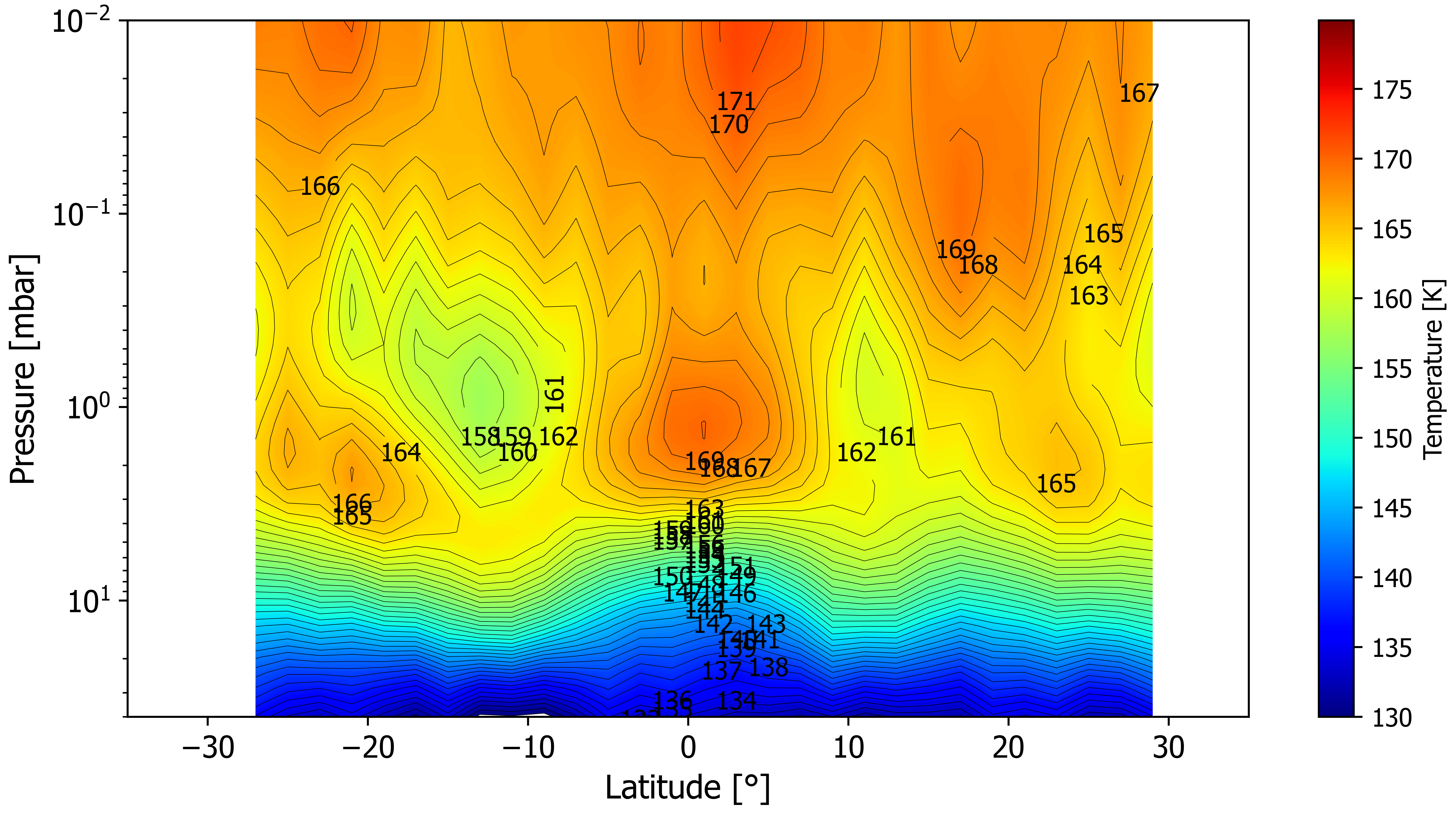}
		\caption{Western limb temperature field resulting from the average between 155$^{\circ}$ and 160$^{\circ}$ longitudes, as observed with Gemini/TEXES on 14, 16, and 20 March 2017. There is no data between 160$^{\circ}$ and 170$^{\circ}$, preventing thus a full coverage of this limb.}
		\label{fig:temperature_west}
\end{figure}

We also produced the temperature maps from the Gemini/TEXES data for the two longitude ranges covered by the limbs observed with ALMA, after accounting for the 15$^{\circ}$ longitudinal smearing of these observations. These temperature maps are shown in Figures \ref{fig:temperature_east} and \ref{fig:temperature_west}. The corresponding wind velocity maps are presented in Figure \ref{fig:wind_map_east_west}. 

\section{\\Longitudinal variability of the temperatures in March 2017 \label{appendix:tempdiffs}}

\begin{figure}[ht]
	\centering
	\includegraphics[width=9cm, keepaspectratio]{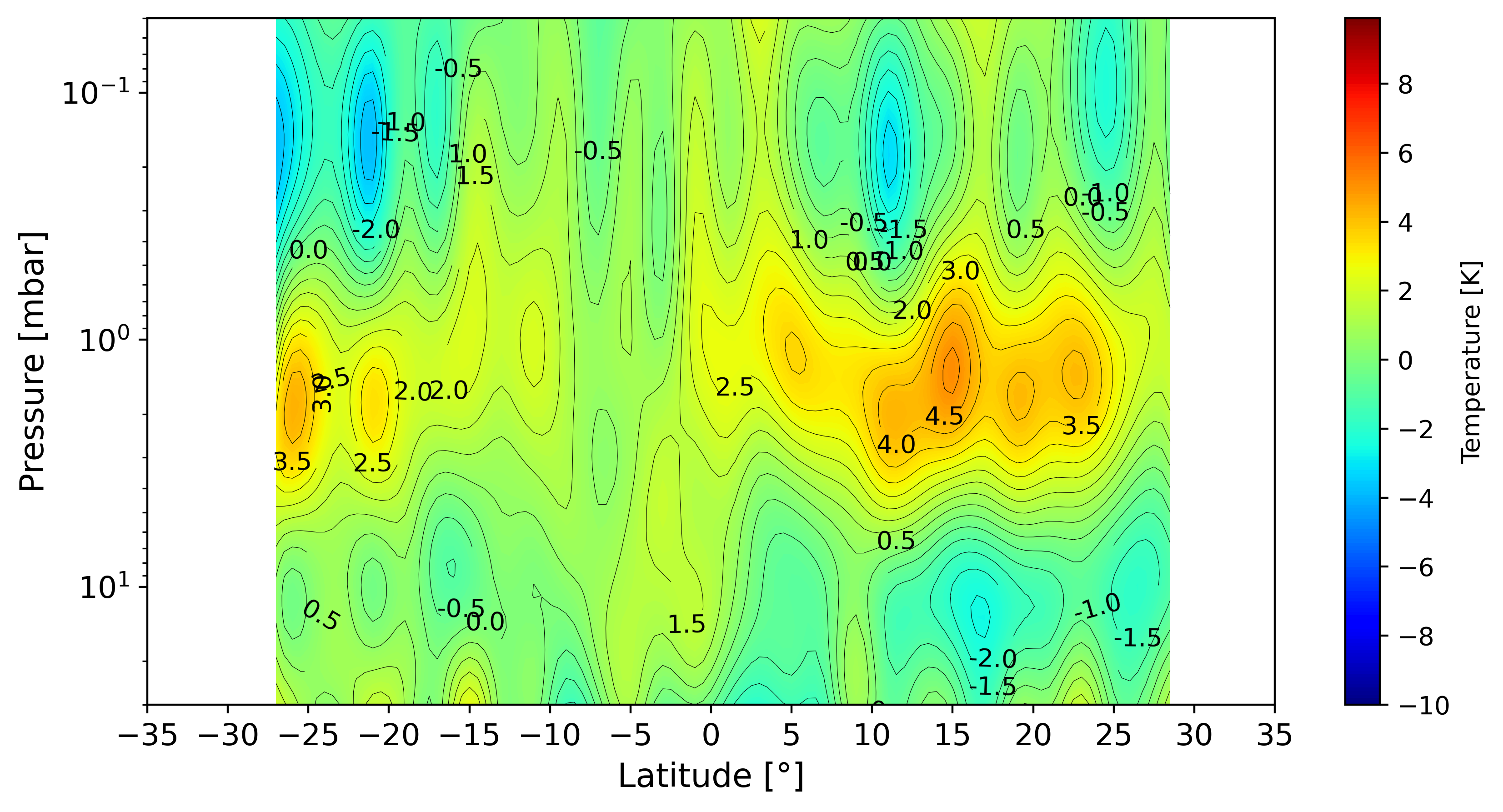}
	\caption{Difference between western limb temperatures and the zonal mean. 
	}
	\label{fig:west-zonal_temperature}
\end{figure}

\begin{figure}[ht]
	\centering
	\includegraphics[width=9cm, keepaspectratio]{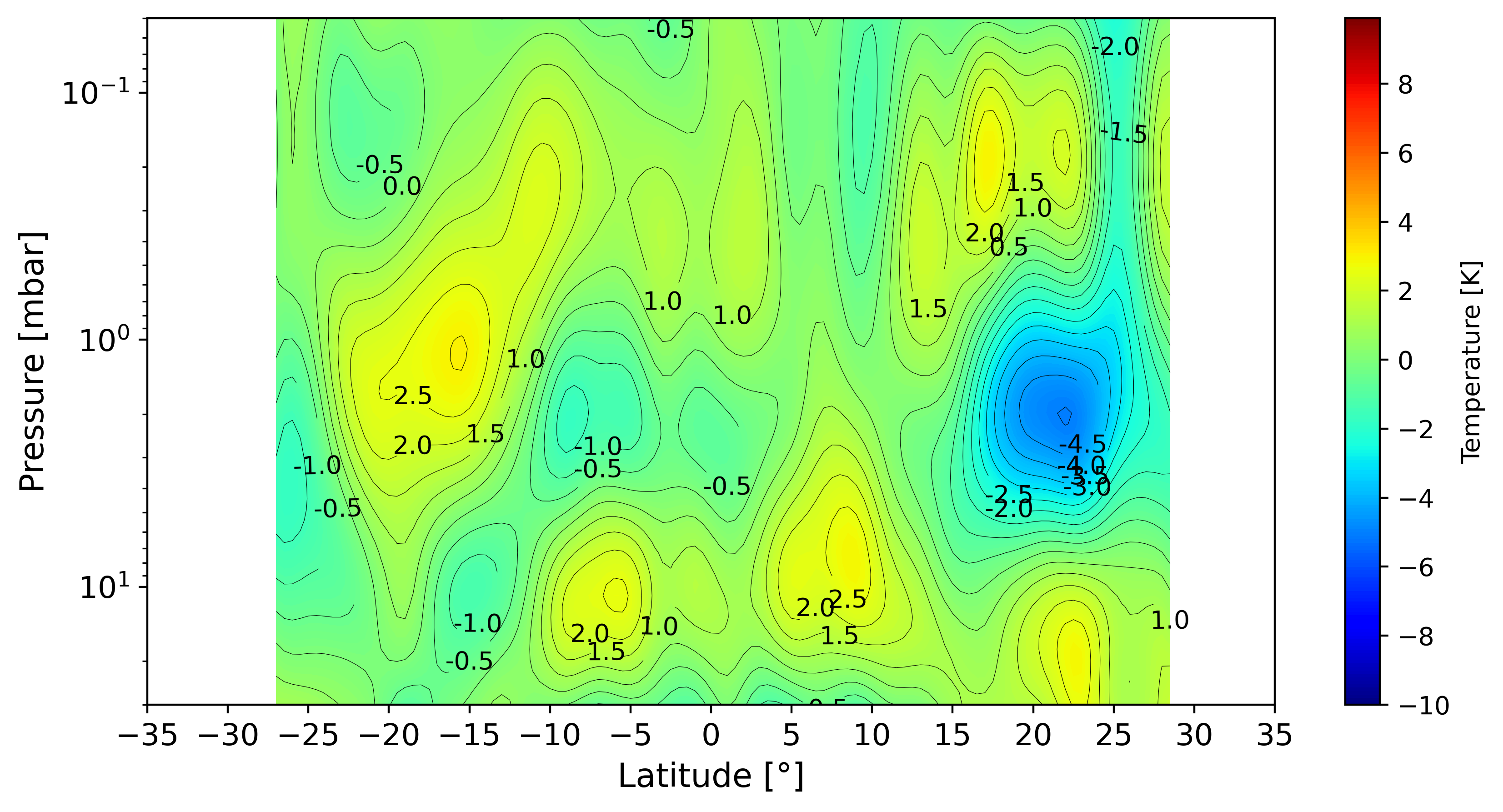}
	\caption{Difference between eastern limb temperatures and the zonal mean. 
	}
	\label{fig:east-zonal_temperature}
\end{figure}

We computed the difference between the Gemini/TEXES temperatures averaged over the western limb and the zonal mean, and proceeded similarly for the eastern limb. The results are shown in Figures \ref{fig:west-zonal_temperature} and \ref{fig:east-zonal_temperature}. 

We find that the western limb presents a hot spot centered at 15$^\circ$N and extended over 10-15$\circ$ in latitude between 1 and 2\,mbar. This hot spot may explain, in part, the differences in wind speeds notably observed around 10$^\circ$N in the two limbs with ALMA (Figure \ref{fig:wind_east-west}). Anticyclonic motions about this spot would decrease the wind speeds at 10$^\circ$N, and increase the wind speeds at 20$^\circ$N on the western limb with respect to the zonal average. On the contrary, the eastern limb shows a cold spot at 2\,mbar and centered at 22$^\circ$N. Here, cyclonic motions about this cold spot would tend to increase wind speeds at $\sim$17$^\circ$N, and decrease them at $\sim$26$^\circ$N on the eastern limb with respect to the zonal average. This would qualitatively tend to bring back in agreement the wind speed profiles from the two limbs of Figure \ref{fig:wind_east-west}, mostly regarding the differences seen on the 10$\circ$N prograde jet. 


\section{\\Wind and temperature data smoothing method \label{appendix:smoothing}}

\begin{figure}[ht] 
	\centering
	\includegraphics[width=9cm, keepaspectratio]{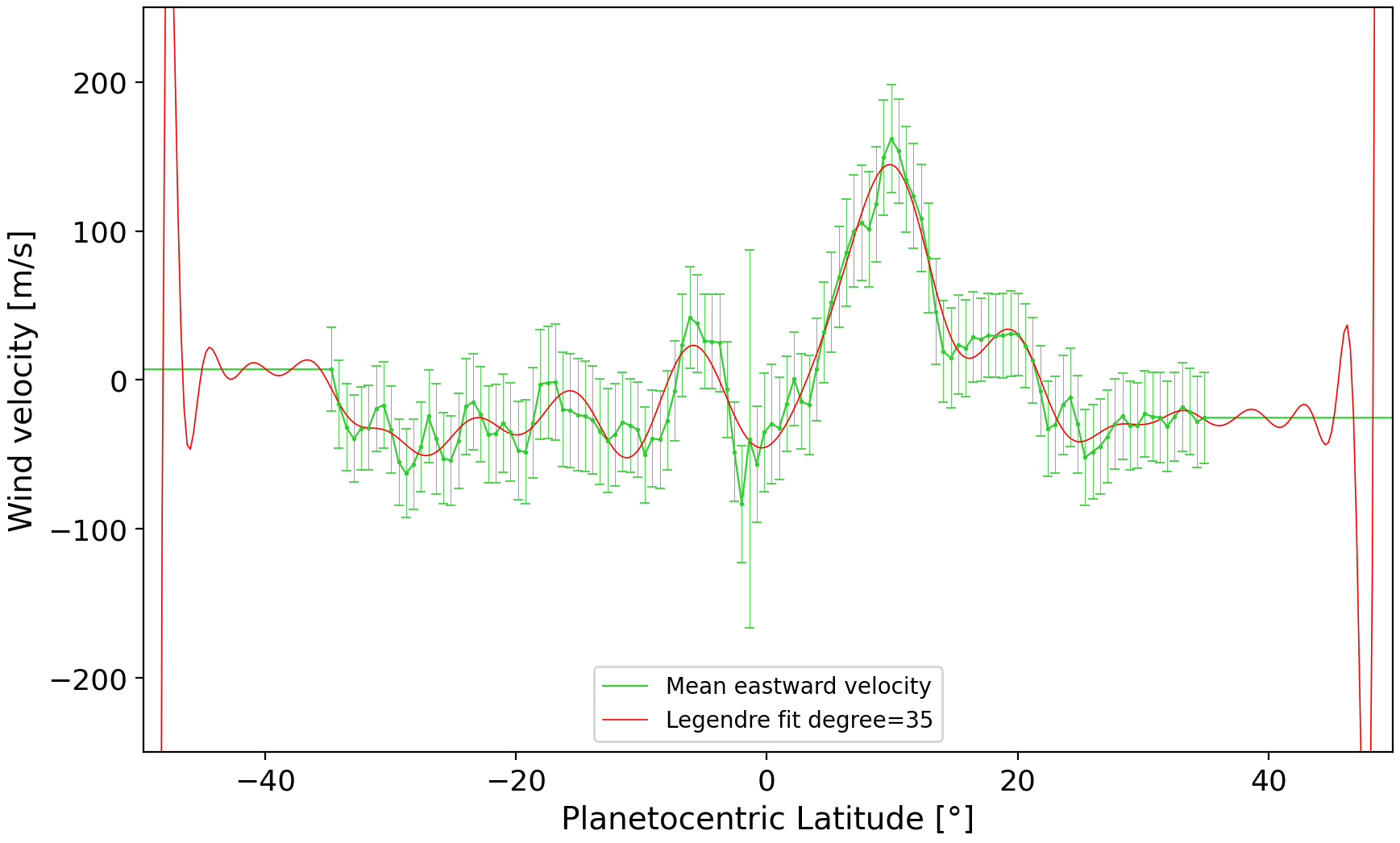}
	\caption{Legendre polynomial series smoothing of the ALMA wind speeds. We extend the fitting range from $\pm$35$^\circ$ to $\pm$50$^\circ$ to limit the edge effects of such fitting procedure to outside the studied interval. The resulting fit is then truncated to the interval of interest.}
	\label{fig:wind_exemple_fit_legendre}
\end{figure}

Before using the wind speeds and temperatures that come from the observations in our modeling, we smoothed the data with Legendre polynomial series. We first determined the order of the highest order $n$ of the series to smooth our data, such that the fit lies within all uncertainties. For the wind speeds of Figure \ref{fig:wind_fited}, we set $n=35$. For the temperature as a function of latitude (and for each altitude), we set $n=17$. Such polynomials can result in edge effects like the Gibbs phenomenon. To avoid this effect, we had to extrapolate the velocity and temperature curves beyond the latitude range we used in our modeling (i.e., from -35$^{\circ}$ to +35$^{\circ}$). We extended the latitudinal range up to +/-50$^{\circ}$ and applied the fit. The results regarding the wind speeds can be found in Figure \ref{fig:wind_exemple_fit_legendre}, where the Gibbs-like effect can be seen around +/-50$^{\circ}$. This effect is thus avoided in the final latitude range we use in our work.

\end{document}